\documentclass[aps,prd,onecolumn,groupedaddress,showpacs,nofootinbib,amssymb]{revtex4}
\usepackage[dvips]{graphicx}
\usepackage{amssymb}
\usepackage{diagbox, float}
\usepackage{amsmath}
\usepackage{graphicx,color}
\usepackage{amsfonts}
\usepackage{bm}
\usepackage{cancel}
\usepackage{comment}
\usepackage{longtable}
\usepackage{multirow}
\usepackage{xcolor}

\usepackage[T2A]{fontenc}

\newcommand{\be}{\begin{equation}}
\newcommand{\ee}{\end{equation}}
\newcommand{\ba}{\begin{eqnarray}}
\newcommand{\ea}{\end{eqnarray}}

\allowdisplaybreaks[4]

\begin{document}

\title{Tsallis Holographic Dark Energy with varying Gravitational Constant vs $\Lambda$CDM : A Comparative Outlook}
\author{Artyom V. Astashenok}
\email{artyom.art@gmail.com}
\author{Alexander S. Tepliakov}
\affiliation{I. Kant Baltic Federal University\\
Laboratory of Astrophysics\\
Nevskogo, 14, Kaliningrad, 236041, Russia\\}%

\tolerance=5000

\begin{abstract}

We investigate the cosmological evolution of a flat Universe filled with Tsallis holographic dark energy (THDE) under the assumption that the gravitational constant $G$ varies with time according to the power-law parametrisation $d\ln G / d\ln a = \alpha_G = \mathrm{const}$. In this framework, the holographic dark energy density is given by $\rho_{\mathrm{DE}} \propto L^{2\gamma - 4}$, where $L$ is the future event horizon and $\gamma$ is the Tsallis non-additivity parameter. The viability of the model is tested against a combination of observational data, including the Pantheon+ Type Ia supernova sample, direct Hubble parameter measurements $H(z)$, and baryon acoustic oscillation (BAO) data. Using Markov Chain Monte Carlo (MCMC) methods and Bayesian analysis tools such as the Deviance Information Criterion (DIC), the Bayes factor, and the suspiciousness metric, we perform a rigorous statistical comparison with the standard $\Lambda$CDM model. Our results show that models with a varying gravitational constant are preferred over $\Lambda$CDM. The best fit to the combined data is obtained for the THDE model with $C < 1$. Furthermore, we analyse the internal consistency of the data sets using the Index of Inconsistency, the $Q_{\mathrm{DMAP}}$ statistic, and Bayesian suspiciousness. We find that the THDE model with varying $G$ and $C < 1$ partially alleviates the tensions between different probes. Our findings suggest that the combination of Tsallis holographic dark energy with a time-varying gravitational constant offers a phenomenologically rich and statistically competitive alternative to the standard cosmological model, though further studies are needed to fully resolve the observational tensions.

\end{abstract}

\pacs{04.50.Kd, 95.36.+x}

\maketitle

\section{Introduction}
 
One of the central challenges of modern theoretical cosmology is interpreting the observed accelerated expansion of the Universe, confirmed by data on Ia supernovae \cite{1}, \cite{2}. Analysis of the luminosity--redshift relationship reveals systematic deviations from the predictions of a cosmological model that accounts only for matter (baryonic and cold dark matter) and radiation. Restoring consistency between theory and observation requires the introduction of a new physical entity---dark energy---which possesses negative pressure and a high degree of spatial homogeneity. Current observational constraints \cite{Amanullah,Blake,LCDM-1,LCDM-2,LCDM-3,LCDM-4,LCDM-5,LCDM-6,LCDM-7}, largely offering valuable evidence yielding our understanding, indicate that the best description of cosmological evolution is provided by the $\Lambda$CDM model, where the role of dark energy is played by the cosmological constant $\Lambda$. Nevertheless, this parameterisation gives rise to fundamental theoretical problems: firstly, the discrepancy between quantum field theory estimates of the vacuum energy and the observed value of $\Lambda$ is of approximately 120 orders of magnitude; secondly, the proximity of the present-day densities of dark energy and matter, known as the cosmic coincidence problem, remains unexplained.

These difficulties have motivated the search for alternative descriptions of the nature of dark energy. Within the quintessence approach, dark energy is modelled as a dynamical scalar field with the equation of state $-1 < w < -1/3$ \cite{Caldwell,Steinhardt,Ferramacho,Caldwell-2}. This framework permits temporal evolution of the energy density and enables the construction of tracking solutions that alleviate the coincidence problem. In parallel, modified gravity theories have been developed in which accelerated expansion arises from nonlinear corrections to the Einstein--Hilbert action, extending beyond the standard geometric interpretation via scalar curvature \cite{Capozziello,Odintsov,Turner}. 

{A further, physically well-motivated approach is based on the holographic principle, which arises from black hole thermodynamics and posits that the number of degrees of freedom in a spatial region is proportional to its boundary area rather than its volume \cite{3,4,5}. When applied to cosmology, this principle leads to the holographic dark energy (HDE) scenario, in which the density of dark energy is bounded from above by the area of the cosmological horizon, specifically by the inverse square of a characteristic infrared cutoff length $L$, i.e., $\rho_\mathrm{DE} \propto M_P^2 L^{-2}$, where $M_P$ is the reduced Planck mass \cite{Wang}. The choice of the infrared cutoff is a crucial aspect of the model. Common alternatives include the Hubble radius $H^{-1}$, the particle horizon, and the future event horizon. It has been shown \cite{Li} that only the choice of the event horizon leads to an equation of state capable of driving the current accelerated expansion and satisfying the observational constraints in a natural manner. Although the use of the future event horizon may appear to introduce a nonlocal dependence on the future evolution of the Universe, such a formulation is physically admissible and has deep connections with the thermodynamics of causal horizons in general relativity \cite{Novik}, \cite{Padmanabhan}.}

{A further generalisation of the holographic dark energy paradigm was proposed in the framework of Tsallis non-extensive thermodynamics \cite{Tsallis-2}, \cite{Tsallis}. Motivated by the fact that gravitational systems are typically characterised by long-range forces and non-Gaussian fluctuations, Tsallis and Cirto introduced a modified entropy expression of the form $S_\gamma \propto A^\gamma$, where $A$ is the area of the horizon and $\gamma$ is a parameter quantifying the degree of non-additivity. This deformation of the standard Bekenstein--Hawking entropy (which corresponds to $\gamma = 1$) gives rise to a modified holographic energy density $\rho_\mathrm{DE} \propto L^{2\gamma - 4}$, thereby significantly altering the cosmological dynamics \cite{Tavayef,Jahromi,Nojiri}. The Tsallis holographic dark energy (THDE) model has been studied extensively in various backgrounds, including non-flat universes, Brans--Dicke gravity, and interacting scenarios \cite{Jahromi,Nojiri,Saridakis}. In particular, it has been shown that for a certain range of the non-additivity parameter $\gamma$, the THDE model can reproduce the standard $\Lambda$CDM behaviour as a limiting case, while for other values it yields a rich phenomenology, including future singularities or transient phantom behaviour \cite{Astashenok1}, \cite{Astashenok2}.}

{An important and physically well-justified extension of cosmological models involves allowing the gravitational constant $G$ to vary with time. This idea has a long history, originally motivated by Dirac's Large Numbers Hypothesis and subsequently developed in the context of scalar--tensor theories of gravity, such as the Brans--Dicke theory \cite{Brans,Dicke,Dirac}. In these theories, $G$ is effectively promoted to a dynamical field, whose evolution is governed by the underlying scalar degree of freedom. Observational bounds on the present-day variation of $G$ are tight, typically $|\dot{G}/G| < 10^{-13}$ yr$^{-1}$, as derived from lunar laser ranging and pulsar timing measurements \cite{Williams}. Nevertheless, even a small variation of $G$ at earlier epochs can have significant effects on the background cosmological evolution and on the formation of large-scale structure. In the context of dark energy, a time-dependent $G$ can alter the Friedmann equations and the conservation laws, thereby modifying the dynamics of the holographic dark energy component in a non-trivial way \cite{Grande}, \cite{Guberina}. In this paper, we shall assume a simple phenomenological law for the variation of $G$ with the scale factor $a$, namely $d\ln G / d\ln a = \alpha_G = \mathrm{const}$, where $\alpha_G$ is a constant parameter. This parametrisation, while empirical, is sufficiently general to capture the leading-order effects and is compatible with a wide range of scalar--tensor models in the slow-roll regime.}

{A further aspect that must be considered in any viable dark energy model is its dynamical stability against small perturbations. The standard stability analysis via the adiabatic squared speed of sound, $c_s^2 = \partial p/\partial \rho$, is not directly applicable to holographic dark energy because the definition of $\rho_\mathrm{DE}$ involves the global future event horizon, thereby introducing a nonlocal dependence on the cosmological evolution. To address this issue, a rigorous stability analysis can be performed by studying scalar perturbations of the infrared cutoff scale $L$ at the cosmological boundary, as demonstrated in \cite{Astashenok2}. This analysis reveals the parameter regions for which the THDE model is classically stable, thereby providing important consistency conditions that can be confronted with observations.}

{Another important direction is the possibility of an energy exchange between dark energy and dark matter. Such a non-minimal coupling, often referred to as the interacting dark energy scenario, is naturally motivated by the fact that the dark sector of the Universe remains largely unknown, and no symmetry principle forbids such an interaction \cite{Wang2}, \cite{DelCampo}. In the interacting framework, the conservation equations for the separate components are modified by a source term $Q$, typically chosen to be proportional to the dark energy density and the Hubble parameter, i.e., $Q \propto H \rho_\mathrm{DE}$. This interaction leads to a dynamical equilibrium between the two components, allowing for a gradual transition to a quasi-de Sitter regime and potentially preventing future singularities \cite{Qihong}, \cite{Astashenok3}. Moreover, it has been demonstrated that an appropriately chosen interaction can induce an effective phantom equation of state ($w < -1$) without the pathological future singularity known as the Big Rip \cite{Astashenok3}. A detailed dynamical systems analysis of such interacting THDE models, including the full phase-space structure and the identification of critical points, was carried out in \cite{Astashenok4}. The results indicate that the fractional density of dark energy can either approach unity or tend to a constant value less than unity, depending on the interaction parameters, thereby offering a new dynamical mechanism for addressing the cosmic coincidence problem.}

{Given these theoretical motivations, in this work we investigate the cosmological evolution of a flat Friedmann--Robertson--Walker universe filled with Tsallis holographic dark energy, under the assumption that the gravitational constant $G$ varies according to the simple power-law parametrisation $G(a) \propto a^{\alpha_G}$. Our main goal is to determine the observational viability of this extended model by confronting it with the latest available data sets, including type Ia supernovae (Pantheon sample), direct measurements of the Hubble parameter $H(z)$, and baryon acoustic oscillation (BAO) observations. We perform a detailed statistical analysis using MCMC chains method to constrain the free parameters of the model. We then compare the performance of our model against the standard $\Lambda$CDM model using the  deviance information criterion (DIC), thereby quantifying the relative improvement in the fit to the data. In particular, we find that for a certain range of the parameter $C < 1$, the THDE model with a varying gravitational constant provides a better description of the combined observations than the $\Lambda$CDM model, suggesting that this framework may offer a more complete and phenomenologically richer description of the late-time accelerated expansion. We also examine the problem of data tensions in our THDE model with vaying gravitational constant. For this purpose we use so called index of inconsistency and some another metrics.}

The structure of paper is the following. In the next section we consider the main cosmological equations, describing evolution of the universe with varying $G$. Then we investigate future and past evolution of the universe for various parameters of the model. The analysis of the model from viewpoint of observational data is performed in the Section IV. The Section V is devoted to detailed statistical analysis of the THDE model from viewpoint of consistency and tensions in the various datasets. In conclusion we summarize the main findings of our investigation.

\section{Basic equations}

{We start from the Hilbert-Einstein action for gravitational field with varying gravitational constant $G(t)$:
\begin{equation}
    S = \int \sqrt{-g}\left(\frac{R}{16\pi G(t)} + \mathcal{L}_m\right) d^4 x
\end{equation}
where $R$ is scalar curvatue and $\mathcal{L}_m$ is the lagrangian for matter fields. If we assume for $G(t) = G_0 \psi(t)$ ($G_0$ is the current gravitational constant) we can see that action coincides with action for Brans-Dicke theory with scalar field $\psi(t)$ \cite{BD}. Therefore we can use well-known equations for this theory \cite{Gcorr}.} For a Friedmann Universe with zero curvature, the metric can be written in the following form:
\begin{equation}\label{metric}
    ds^2 = dt^2 - a^2(t)\left(dr^2 + r^2 d\Omega^2\right),
\end{equation}
where $a(t)$ is the scale factor. For the Hubble parameter $H\equiv \dot{a}/{a}$, we obtain the equation
\begin{equation}\label{hubble}
    H^2 = \frac{8\pi G}{3}\rho + H \frac{\dot{G}}{G}.
\end{equation}
{We consider that the value of $G'/G$ is small in the late time universe and therefore we omitted higher derivatives of $G$ and powers of $G'/G$ in Eq. (\ref{hubble}).} For the energy density, we assume that the Universe consists of dark energy and matter, i.e., $\rho = \rho_{\mathrm{de}} + \rho_m$. For $\rho$, the continuity equation is the same as in the case where $G=G_{0}=\mathrm{const}$:
\begin{equation}\label{rho_t}
\dot{\rho} + 3H\left(\rho + \frac{p}{c^2}\right) = 0,
\end{equation}
where $p$ is the pressure of all components of the Universe. {Is the continuity equation satisfied for matter separately from the dark component? This is a subtle point in considered model with varying gravitational constant. We consider the non-interacting scenario between matter and HDE as a baseline model. Since we utilize the future event horizon as the infrared cutoff the density of the HDE is inherently defined 
by the global causal structure of the universe.} 

{Introducing a direct phenomenological interaction between locally 
conserved matter and a fundamentally non-local quantity determined by a future 
integral would introduce severe causality issues. Instead, in our model, 
the time-variation of $G(t)$ alters the expansion rate of the universe, which 
self-consistently modulates the evolution of the future event horizon  
and shifts the effective equation of state $w_{de}$. Therefore, the individual 
conservation of matter and HDE is not merely a simplification, but a physically 
justified choice that respects the causal nature of the holographic framework 
under a dynamic gravitational background.} Therefore, for the matter component, we can write
\begin{equation}\label{rho_m}
\dot{\rho}_{m} + 3H\rho_m = 0.
\end{equation}

{The conventional approach to HDE traces back to the well-known Bekenstein entropy bound. For a spatial domain of characteristic size \(L\), the entropy \(S\) and the dark energy density \(\rho_{de}\) satisfy the inequality}

\begin{equation}
\rho_{de} L^4 \leq S,
\label{eq:bekenstein}
\end{equation}
{with the entropy scaling as \(S \sim L^2\) in the standard picture.}

{Departing from this, Tsallis and Cirto \cite{Tsallis-2} proposed a modified entropy--area relation that incorporates potential quantum corrections, namely}
\begin{equation}
S = \delta S^\gamma = (4\pi)^\gamma \delta L^{2\gamma},
\label{eq:tsallis}
\end{equation}
{where the non-additivity parameter \(\gamma\) need not be equal to unity.}

{In the work by Cohen and collaborators \cite{Cohen}, a relationship was established between the entropy, the infrared cutoff \(L_0\), and the ultraviolet cutoff \(\Lambda\):}
\begin{equation}
L_0 \Lambda \leq S^{1/4}.
\label{eq:cohen}
\end{equation}

{Inserting the Tsallis--Cirto entropy expression (\ref{eq:tsallis}) into the above inequality yields an upper bound on the ultraviolet cutoff:}
\begin{equation}
\Lambda^4 \leq \delta (4\pi)^\gamma L^{2\gamma - 4}.
\label{eq:Lambda_bound}
\end{equation}

{Adopting the standard HDE conjecture, one identifies \(\Lambda^4\) with the dark energy density \(\rho_{de}\). This immediately leads to}
\begin{equation}
\rho_{de} = \frac{3C^2 M_{pl}^2}{L^{4 - 2\gamma}},
\label{eq:rhode}
\end{equation}
{where the coefficient is defined as}
\begin{equation}
C^2 = \frac{\delta (4\pi)^\gamma}{3},
\label{eq:C2}
\end{equation}
{which is an unspecified constant carrying mass dimension $M^{2\gamma}$.}

{In a case of varying gravitational constant} we adopt the following relation for holographic component density:
\begin{equation}
   \rho_{\mathrm{de}} = \frac{3C^2}{8\pi G L^{4-2\gamma}},
\end{equation}
where $L$ is a characteristic scale in the Universe, $\gamma$ is the non-additivity parameter, and $C$ is a fixed constant. For the scale $L$, we use the event horizon length:
$$
L(t)=a(t)\int_{t}^{t_f} \frac{dt'}{a(t')}.
$$
Here, $t_f$ is the time of the final singularity, or $t_f=\infty$ if there is no future singularity. The equation for $L$ is easily obtained by differentiation with respect to time $t$:
\begin{equation}
    \dot{L} = HL - 1.
\end{equation}
We also write the equation for the time derivative of the Hubble parameter:
\begin{equation}\label{H_t}
2\dot{H} = \frac{1}{1 - \frac{\dot{G}}{HG}}\left( \frac{8\pi G}{3}\frac{\dot{G}}{HG}{\rho} + \frac{8\pi G}{3H}\dot{\rho} + H \frac{\mathrm{d}}{\mathrm{d} t}\left(\frac{\dot{G}}{HG}\right) \right).
\end{equation}

It is useful to use the variable $x\equiv \ln a$ instead of time. In terms of this variable, we can write the following system of cosmological equations, assuming that $8\pi G_{0} = c = 1$ (where $G_0$ is the current gravitational constant):

For the matter and dark energy densities:
\begin{equation}\label{rho_x}
\rho_m' = - 3 \rho_m, 
\end{equation}
\begin{equation}\label{rho_d_x}
\rho_{\mathrm{de}}' = \rho_{\mathrm{de}} \left( - \alpha_G + (2\gamma-4) \left( 1 - \frac{1}{LH} \right)\right),
\end{equation}
where $'\equiv d/dx$, $\alpha_G = \tilde{G}'/\tilde{G}$, and $\tilde{G}=G/G_{0}$ is the dimensionless gravitational constant in units of its current value. The equation for the scale $L$ can be rewritten as
\begin{equation}\label{L_x}
    L' = L - \frac{1}{H}.
\end{equation}
Finally, Eq.~(\ref{H_t}) can be rewritten as
\begin{equation}\label{H_x}
2H' = \frac{1}{1 -\alpha_G}\left(\frac{\tilde{G}}{H}\frac{\alpha_{G}\rho}{3} + \frac{\tilde{G}}{3H}\rho' + {H} \alpha_G' \right).
\end{equation}
The evolution of the Universe can be determined by integrating Eqs.~(\ref{rho_x})--(\ref{H_x}) with initial conditions. We start from $a_0 = 1$, which corresponds to $x_0 = 0$. For $H(0)$, we fix the value $H(0) = 1$; therefore, $H$ represents the Hubble parameter in units of $H_0$. At the current moment of time, the Friedmann equation
\begin{equation}
    H^2 = \tilde{G} \frac{\rho}{3} + H^2 \alpha_G
\end{equation}
implies that 
$$(\Omega_{\mathrm{de}} + \Omega_m)_{x=0} = (1 - \alpha_G), \quad \Omega_{m,\mathrm{de}} \equiv  \frac{\rho_{m,\mathrm{de}}}{3H^2}.$$
Therefore, $\Omega_{m}$ and $\Omega_{\mathrm{de}}$ represent the matter and dark energy density fractions, respectively, from the viewpoint of cosmology with a fixed gravitational constant. {One note that sum of dark energy and matter fractions differs from $1$. It is similar to the cosmology with non-zero curvature \cite{Gcorr-2}.}

Finally, for the scale $L$, using the equation for $\rho_{\mathrm{de}}$
\begin{equation}
 \rho_{\mathrm{de}} = \frac{3C^2}{\tilde{G} L^{4-2\gamma}},   
\end{equation}
we obtain
$$
L(0) = \left(\frac{C^2}{\Omega_{\mathrm{de}}(0)}\right)^{\frac{1}{4-2\gamma}}.
$$
It is interesting to investigate the cosmological evolution of the Universe for various values of $C$ and $\gamma$. We consider the simple case where $\alpha_{G}=\mathrm{const}$. {Of course this is a strong ansatz but we can for simplicity fit the possible dependence of $G$ by a single parameter assuming that $\alpha_G$ is small constant. The second question is affect of such a power-law evolution of $G$ on early-universe physics. We agree that such model cannot be considered as suitable model for the distant past and therefore use this approximation for only limit range of redshift ($z<2.5$) assuming some another law for $G$ for an earlier times. }



\section{Cosmological evolution and observational data}
\begin{figure}
    \centering
    \includegraphics[scale=0.28]{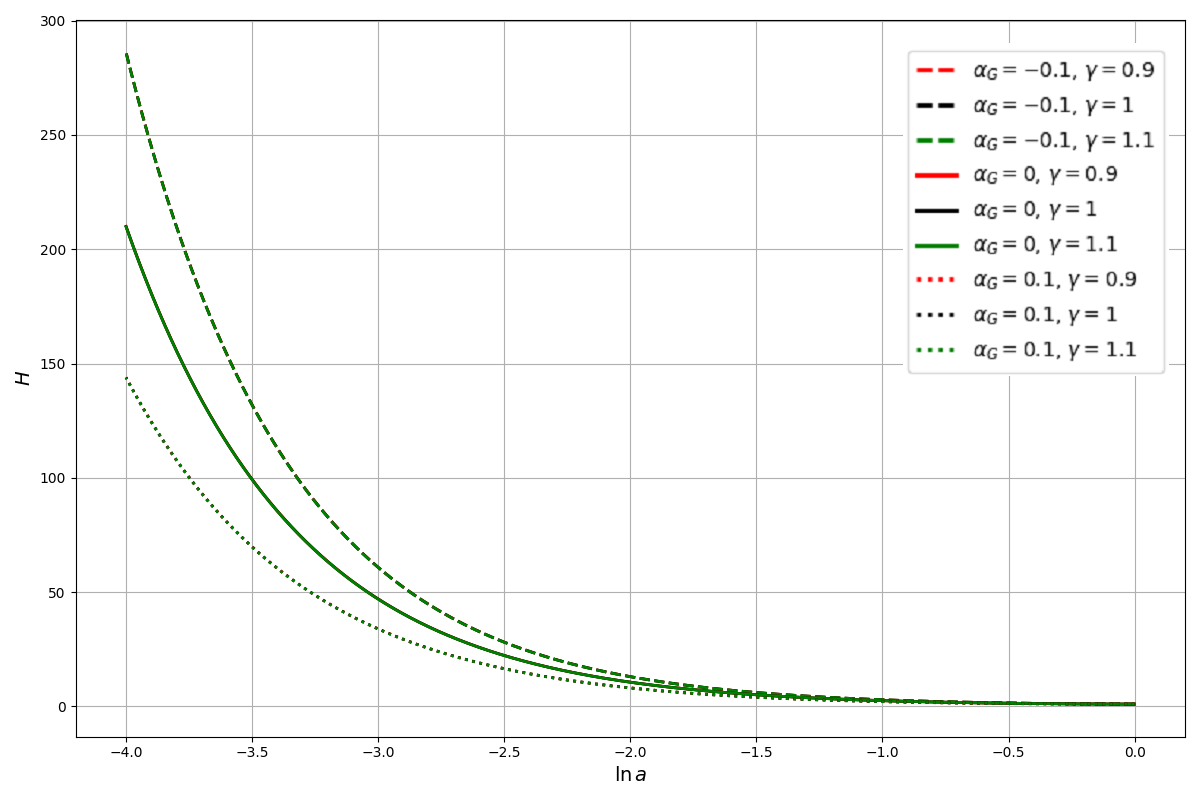}\includegraphics[scale=0.28]{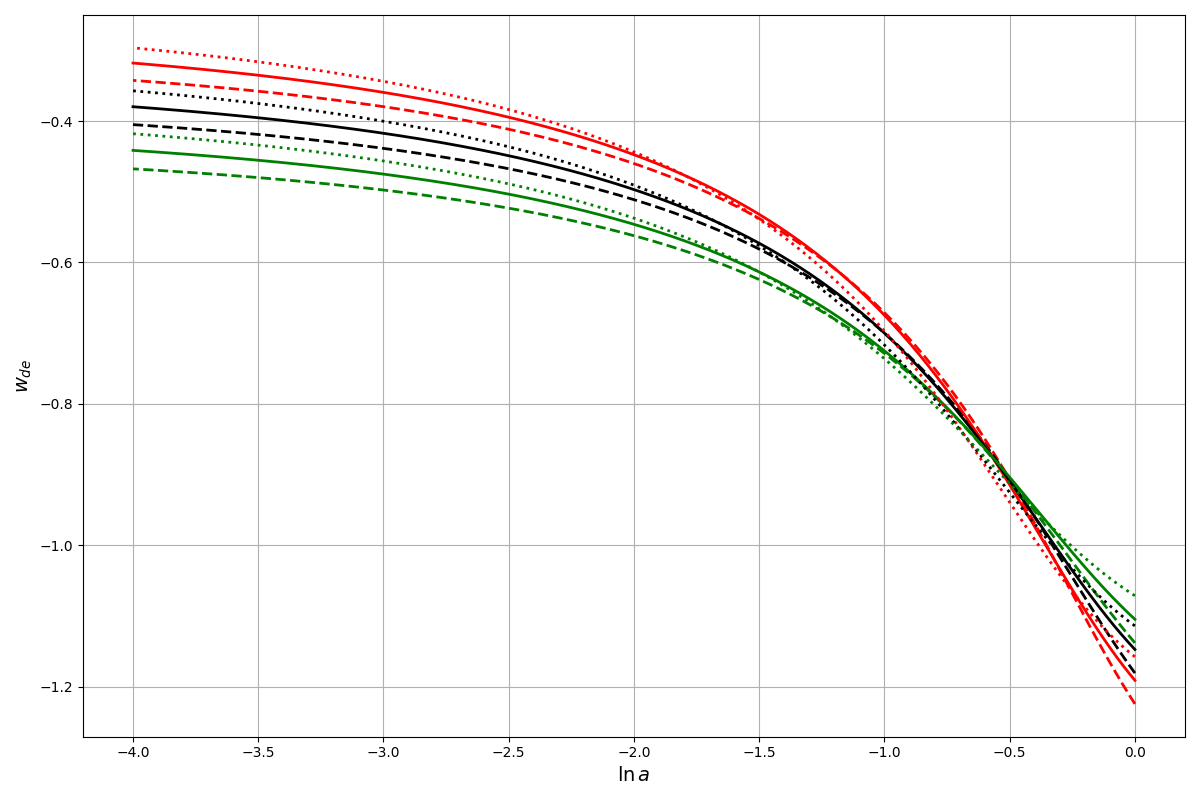}
    \caption{The dependence of the dimensionless Hubble parameter (left panel) and the equation-of-state parameter $w_{de}$ (right panel) on the logarithm of the scale factor for $C=0.7$ in the past, for selected values of $\gamma$ and $\alpha_G$.}
    \label{fig:H1}
\end{figure}

First, we present the evolution of the dimensionless Hubble parameter in the past for $C=0.7$ (see Fig.~\ref{fig:H1}). Values of $\alpha_G = \pm 0.1$ are considered. As an initial condition for $\Omega_{de}$, we adopt the value $\Omega_{de}=0.72$. For a fixed $\Omega_{de}$, the Hubble parameter depends mainly on $\alpha_G$. We observe that the dependence on $\gamma$ is very weak. This is expected, since in the past the cosmological evolution is dominated by the matter component, and the influence of dark energy is negligible.

\begin{figure}
    \centering
    \includegraphics[scale=0.28]{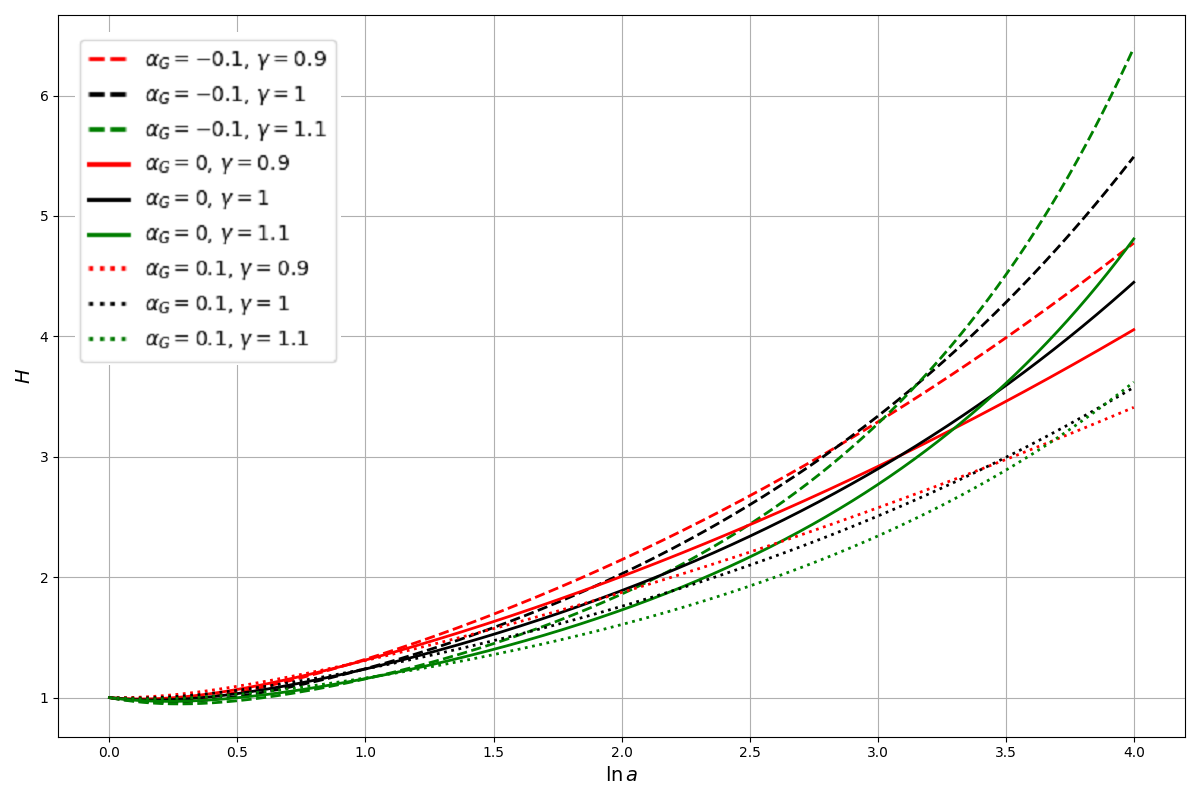}\includegraphics[scale=0.28]{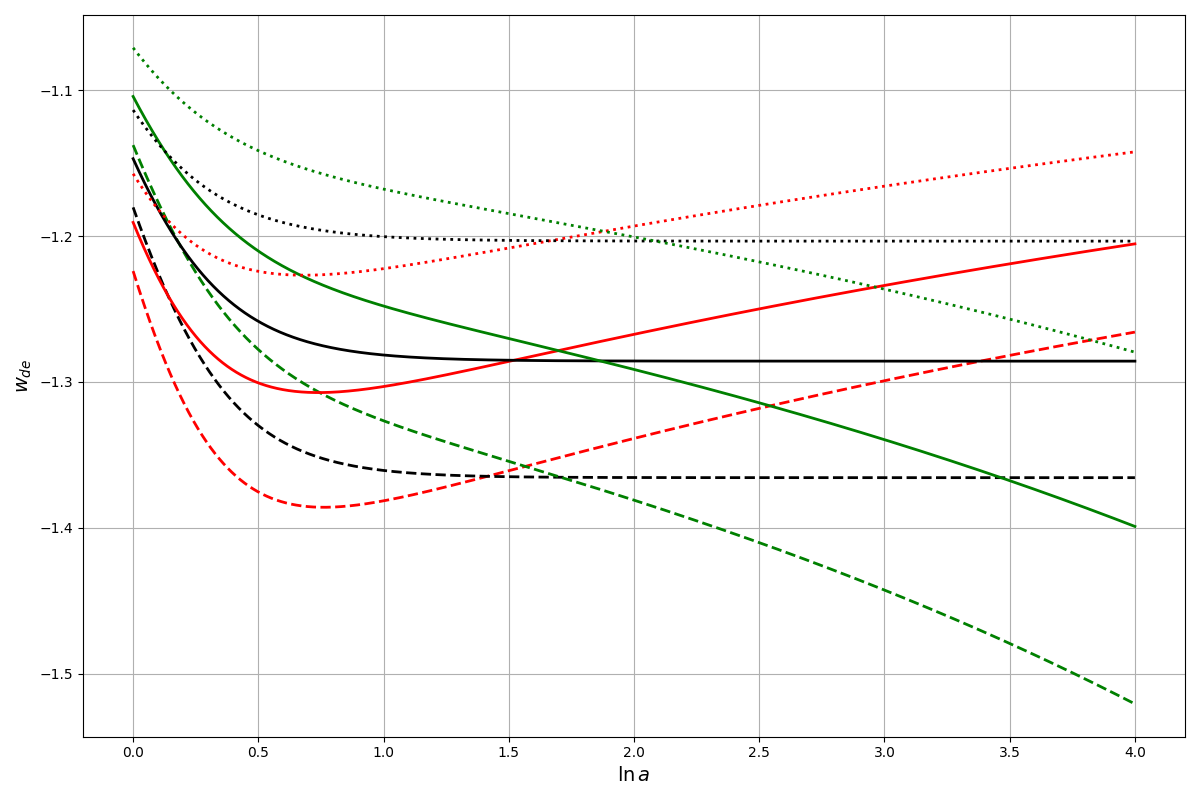}
    \caption{The dependence of the dimensionless Hubble parameter (left panel) and the equation-of-state parameter $w_{de}$ (right panel) on the logarithm of the scale factor for $C=0.7$ in the future, for selected values of $\gamma$ and $\alpha_G$.}
    \label{fig:H2}
\end{figure}

It is also interesting to consider the evolution of the equation-of-state parameter $w_{de}=p_{de}/\rho_{de}$ for holographic dark energy (Fig.~\ref{fig:H1}). For $C=0.7$ and the considered value of $\Omega_{de}$ in the past, phantomization occurs, i.e., $w_{de}$ crosses the phantom divide $w_{de} = -1$. Considering the future cosmological evolution, we observe (Fig.~\ref{fig:H2}) a clear dependence on $\gamma$. Negative values of $\alpha_G$ lead to a more rapid growth of the Hubble parameter with expansion in comparison with the standard model without evolution of $G$.

From the evolution of $w_{de}$, we conclude that for $\gamma \geq 1$, a future singularity occurs.

\begin{figure}
    \centering
    \includegraphics[scale=0.28]{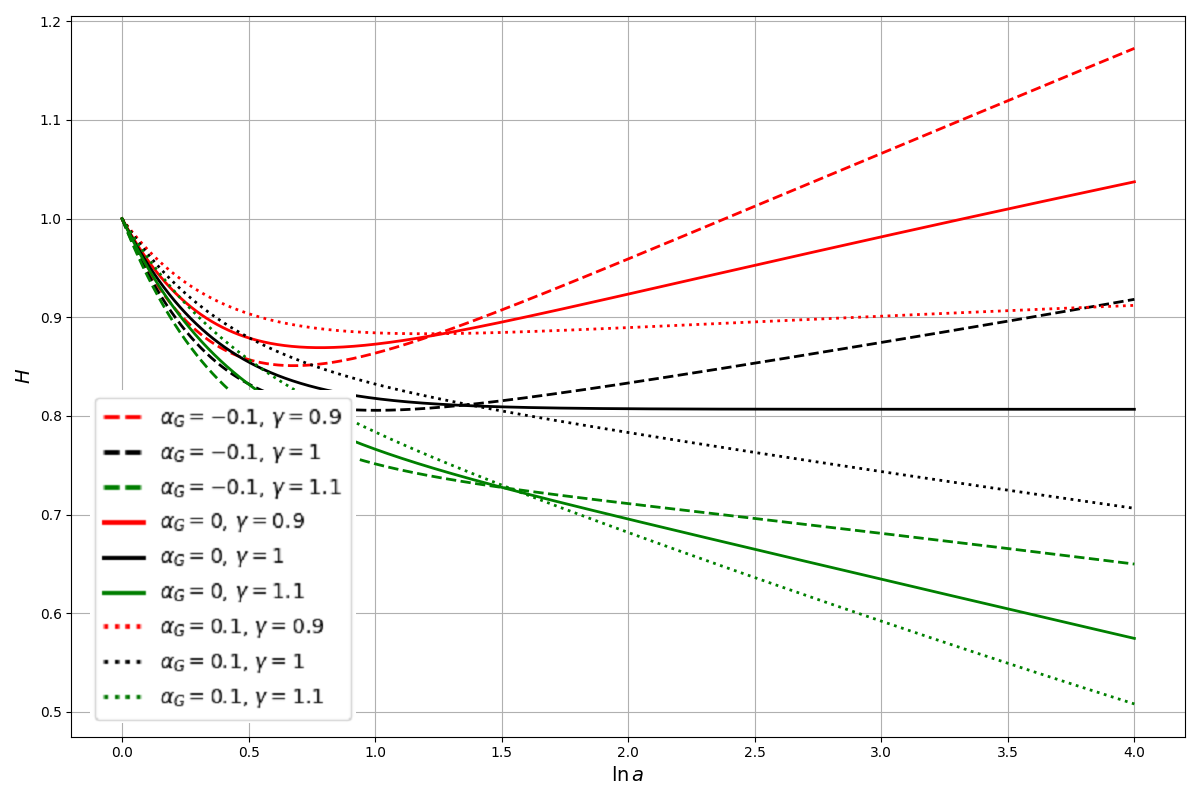}\includegraphics[scale=0.28]{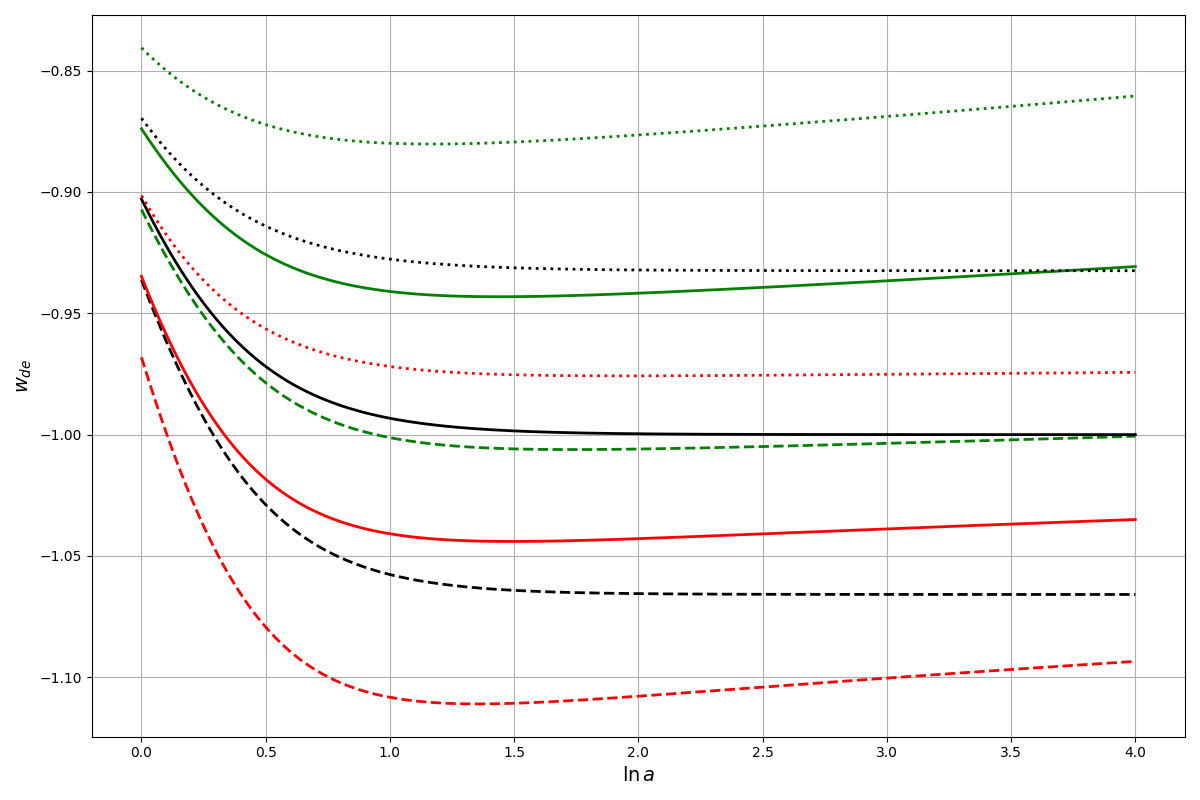}
    \caption{The same as in Fig.~\ref{fig:H2}, but for $C=1.0$ and selected values of $\gamma$ and $\alpha_G$.}
    \label{fig:H3}
\end{figure}

For $C=1$, the analysis shows that phantomization can occur in the future for negative values of $\alpha_G$ (Fig.~\ref{fig:H3}). However, this phantomization does not lead to a future singularity. For example, for $\gamma=1.1$ and $\alpha_G=-0.1$, the Hubble parameter decreases with time, as in other cases with $\gamma=1.1$. For $\gamma\leq 1$ and $\alpha_G=-0.1$, the Hubble parameter grows slowly with time.

\begin{figure}
    \centering
    \includegraphics[scale=0.28]{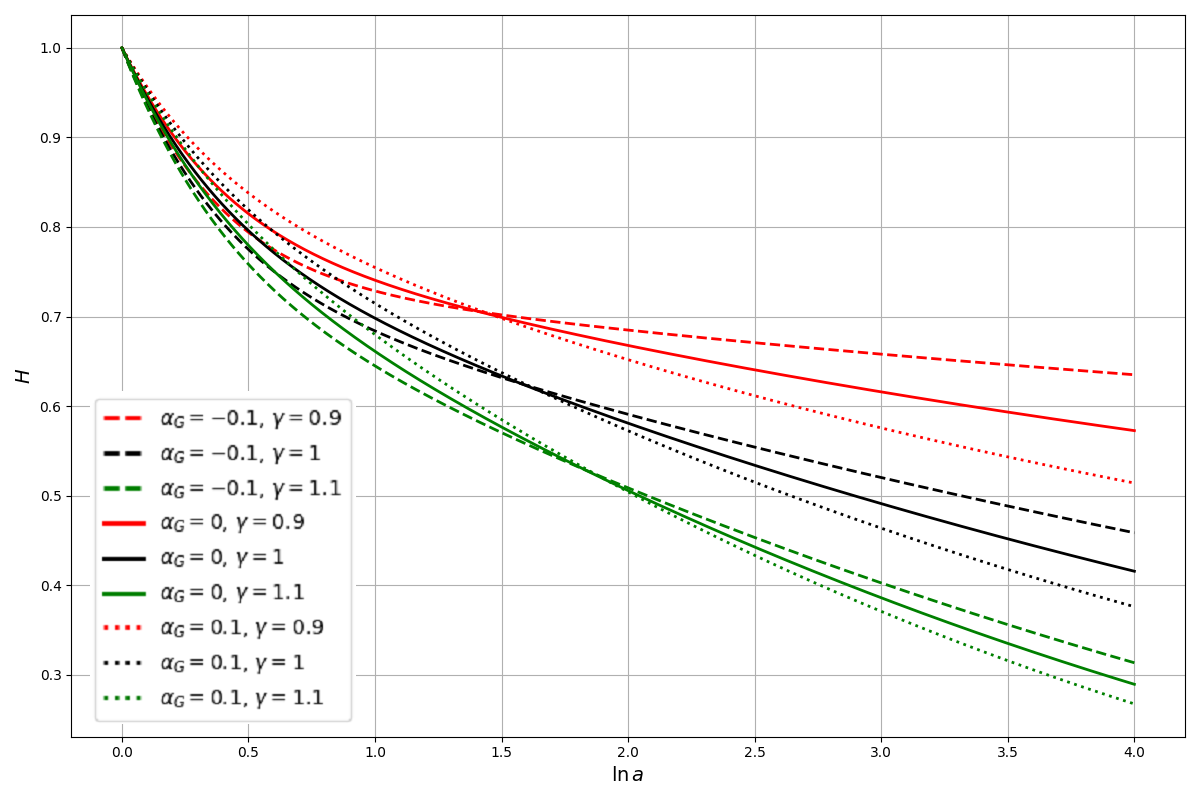}\includegraphics[scale=0.28]{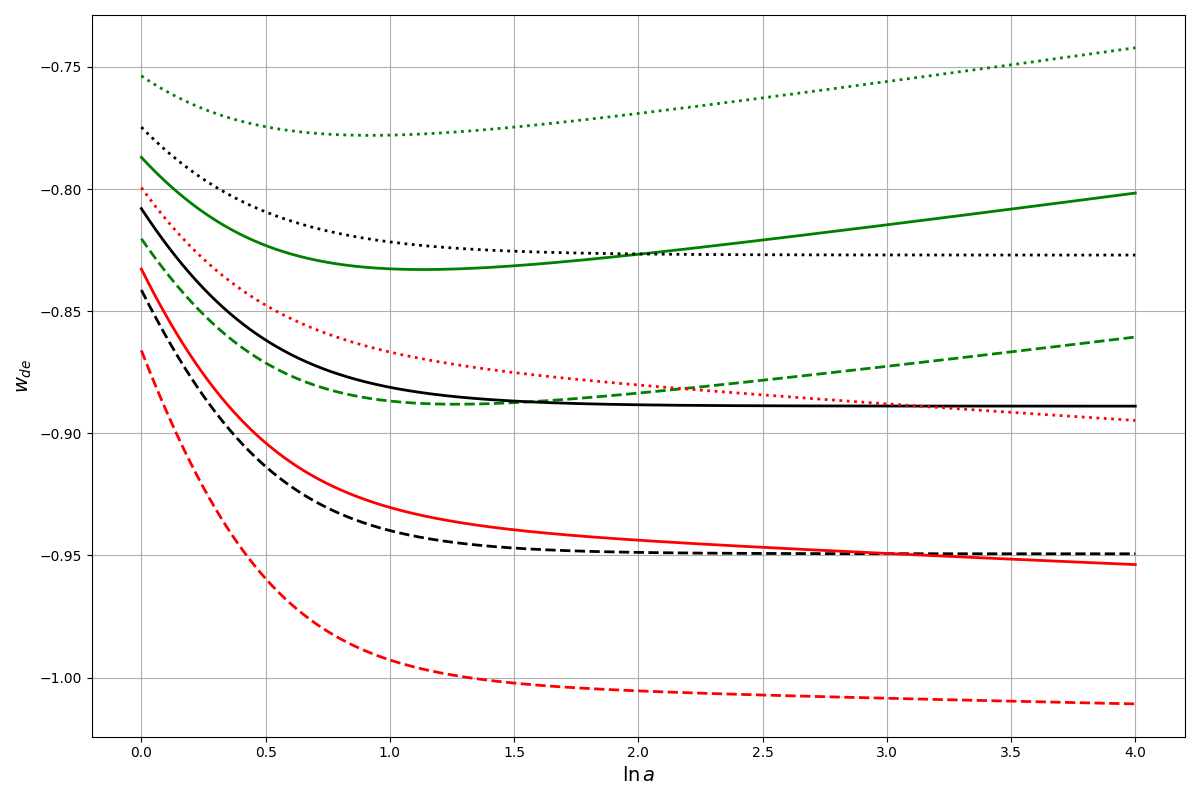}
    \caption{The same as in Fig.~\ref{fig:H2}, but for $C=1.2$ and selected values of $\gamma$ and $\alpha_G$. For a given $\gamma$, the Hubble parameter decreases with time more slowly for positive values of $\alpha_G$.}
    \label{fig:H4}
\end{figure}

For $C=1.2$, phantomization occurs due to the decreasing gravitational constant in the future ($\alpha_G<0$) for $\gamma<1$. In all cases, however, the Hubble parameter decreases (see Fig.~\ref{fig:H4}).

The next step is to consider the correspondence between the THDE model with a varying gravitational constant and observational data. The observational data included in our analysis comprise: (i) the distance modulus--redshift relation for Type Ia supernovae; (ii) measurements of the Hubble parameter as a function of redshift; and (iii) distance indicators from baryon acoustic oscillations (BAO).

\subsection{SNe Ia test}  

We use the Pantheon+ dataset for the Type Ia supernova distance modulus--redshift relation, which consists of 1701 light curves \cite{Torny, Skolnik, Brout}.

The theoretical value of the distance modulus for a given supernova, in the case where the gravitational constant varies, must account for the dependence of the Chandrasekhar mass $M_{\mathrm{Ch}}$ (and therefore the peak luminosity of the supernova) on the effective gravitational constant. According to \cite{Gaztanaga}, the peak luminosity of a supernova is proportional to the mass of synthesized nickel, which is a fixed fraction of $M_{\mathrm{Ch}} \sim G^{-3/2}$. Since the luminosity distance is inversely proportional to the square root of the luminosity, the distance modulus obeys the following relation:
\begin{equation}
    \boldsymbol{\mu} = 5 \log_{10}\left[\frac{d_L(z_{\mathrm{hel}},z_{\mathrm{cmb}})}{\mathrm{Mpc}}\right] + 25 + \frac{15}{4}\log_{10} \tilde{G},
\end{equation}
where $z_{\mathrm{cmb}}$ and $z_{\mathrm{hel}}$ are the cosmic microwave background rest-frame and heliocentric redshifts of the supernova, respectively. The luminosity distance $d_L$ is given by
\begin{equation}
    d_L(z_{\mathrm{hel}},z_{\mathrm{cmb}}) = (1+z_{\mathrm{hel}}) r(z_{\mathrm{cmb}}),
\end{equation}
where the corresponding comoving distance $r(z)$ for a spatially flat Universe is
\begin{equation}
\label{eq:rz}
r(z) = H_0^{-1} \int_0^z\frac{dz'}{E(z')},
\end{equation}
and $E(z) \equiv H(z)/H_0$ is the dimensionless Hubble parameter.

The optimal model parameters correspond to the minimum of the $\chi^2$ criterion:
$$
\chi^{2}_{\mathrm{SNe}} = \Delta \boldsymbol{\mu}^T \cdot \mathbf{C}^{-1}_{\mathrm{SNe}} \cdot \Delta\boldsymbol{\mu},
$$
where the $i$-th component of the vector $\Delta \boldsymbol{\mu}$ is
$$
\Delta\mu_{i}\equiv \mu_{i}-\mu(z_{i,\mathrm{hel}}, z_{i,\mathrm{cmb}}).
$$
The total covariance matrix $\mathbf{C}_{\mathrm{SNe}}$ consists of statistical and systematic components:
\begin{equation}
\mathbf{C}_{\mathrm{SNe}}=\mathbf{D}_{\mathrm{stat}}+\mathbf{C}_{\mathrm{stat}} +\mathbf{C}_{\mathrm{sys}}.
\end{equation}
Here, $\mathbf{D}_{\mathrm{stat}}$ is the diagonal part of the statistical uncertainty, as given in \cite{Skolnik}, while $\mathbf{C}_{\mathrm{stat}}$ and $\mathbf{C}_{\mathrm{sys}}$ are the statistical and systematic covariance matrices, respectively, which are provided in Ref.~\cite{Pantheon_SH0}. The uncertainties in the distance modulus values are also tabulated in that source.

Let us introduce the vector
$$
\boldsymbol{\mu}^{(0)} = \boldsymbol{\mu} - \mu_{0}, \quad \mu_0 =  5\log_{10}\left(\frac{d_{H_{0}}}{\mathrm{Mpc}}\right),
$$
where $d_{H_{0}} = c/H_0$. We can determine the optimal value of the current Hubble parameter by minimizing the $\chi^2$ function for the SNe Ia data, which can be written in the form
\begin{equation}  \label{eq:chi2_SN}
    \chi^2_{\mathrm{SNe}} = A_{\mathrm{SNe}} - 2 B_{\mathrm{SNe}} \mu_0  + S_{\mathrm{SNe}}\mu_{0}^{2},
\end{equation}
where the coefficients $A_{\mathrm{SNe}}$, $B_{\mathrm{SNe}}$, and $S_{\mathrm{SNe}}$ do not depend on $H_0$:
$$
A_{\mathrm{SNe}} = (\boldsymbol{\mu}_{\mathrm{obs}}-\boldsymbol{\mu}^{(0)})^{T}\cdot \mathbf{C}^{-1}_{\mathrm{SNe}} \cdot (\boldsymbol{\mu}_{\mathrm{obs}}-\boldsymbol{\mu}^{(0)}),
$$
$$
B_{\mathrm{SNe}} = (\boldsymbol{\mu}_{\mathrm{obs}} - \boldsymbol{\mu}^{(0)})^{T}\cdot \mathbf{C}^{-1}_{\mathrm{SNe}} \mathbf{I}, \quad S_{\mathrm{SNe}} = \mathbf{I}^T\mathbf{C}^{-1}_{\mathrm{SNe}}\mathbf{I}.
$$
Here $\mathbf{I}^T = (1,1,\dots,1)$. The function $\chi^2_{\mathrm{SNe}}$ attains its minimum at
$$
\mu_{0}^{(\mathrm{opt})} = B_{\mathrm{SNe}}/S_{\mathrm{SNe}}.
$$
Thus, we can determine the best-fit value of $H_0$ from the SNe Ia data.

\subsection{BAO test}

The BAO measurements \cite{Blake} include several distinct observables, namely: the ratio of the Hubble distance to the sound horizon at the drag epoch $r_d$,
$$
\frac{d_H(z)}{r_d}=\frac{c\,r_d^{-1}}{H(z)};
$$
the distance parameter $d_{V}(z)$,
$$
\frac{d_V(z)}{r_d}= \left[\frac{c\,z\,r_d^{-3} d_L^2(z)}{H(z)(1+z)^2}\right]^{\frac{1}{3}},
$$
where $c$ is the speed of sound in the epoch of recombination; and, finally, the ratio $r(z)/r_d$.

For these quantities, we use data from \cite{DESI:2024mwx}, which were obtained for the redshift range $0.1<z<2.4$ and split into 7 distinct redshift bins (see Table~\ref{tab:BAO}). It should be noted that these measurements are effectively independent of each other; thus, no covariance matrix is considered.

BAO measurements are insensitive to the value of the product $H_0 r_d$. In this analysis, rather than fixing the sound horizon at the drag epoch (\(r_{d}\)) to a model-dependent value or a point estimate, we treat \(r_{d}\) as a free nuisance parameter with an uninformative flat prior bounded by \([140, 155]\) Mpc. This specific range is well-motivated by both early- and late-universe cosmological considerations for the following reasons: inclusion of early dark energy and new physics. To alleviate the persistent $H_0$ tension between early- and late-universe probes, various extensions to the standard $\Lambda$CDM model have been proposed. Prominent solutions, such as early dark energy  or additional effective relativistic degrees of freedom (\(\Delta N_{\text{eff}}\)), systematically shift the sound horizon to lower values. Lowering our prior boundary to 135 Mpc ensures that our analysis remains fully agnostic to such early-universe modifications and avoids artificially penalizing models that resolve the Hubble tension.

\begin{table}[ht]
\centering
\setlength{\tabcolsep}{1.em}
\renewcommand{\arraystretch}{1.1}
\begin{tabular}{|l|c|c|c|c|}
\hline
Tracer  & $z_{\mathrm{eff}}$ & $d_H/r_d$         & $d_V/r_d$     & $r/r_d$ \\
\hline
BGS     & $0.30$        & $-$               & $7.93\pm0.15$ & \\
LRG     & $0.51$        & $20.98\pm0.61$    & $-$           & $13.62\pm 0.25$\\
LRG     & $0.71$        & $20.08\pm0.60$    & $-$           & $16.85\pm 0.32$\\
LRG+ELG & $0.93$        & $17.88\pm0.35$    & $-$           & $21.71\pm 0.28$ \\
ELG     & $1.32$        & $13.82\pm0.42$    & $-$           & $27.79\pm 0.69$ \\
QSO     & $1.49$        & $-$               & $26.07\pm0.67$ & \\
Lya QSO & $2.33$        & $8.52\pm0.17$     & $-$            & $39.71\pm 0.94$ \\
\hline
\end{tabular}
\caption{DESI-BAO data with tracers, effective redshifts $z_{\mathrm{eff}}$, and ratios $d_H/r_d$, $d_V/r_d$, and $r/r_d$ \cite{DESI:2024mwx}.}
\label{tab:BAO}
\end{table}

For all three types of BAO measurements, denoted by $X_i$, we calculate
\begin{equation}
\label{loglikeBAOu}
    \chi^{2}_{\mathrm{BAO}} = \sum_{i=1}^{N_{\mathrm{X}}}\left[\dfrac{X_i-X(z_i)}{\sigma_{X_i}}\right]^2.
\end{equation}
Alternatively, this expression can be rewritten as
\begin{equation}
   \chi^{2}_{\mathrm{BAO}} = A_{\mathrm{BAO}} - 2B_{\mathrm{BAO}}\nu + \nu^2 S_{\mathrm{BAO}}, \quad \nu = \frac{1}{r_d H_0},
\end{equation}
where
$$
A_{\mathrm{BAO}} = \sum_{i=1}^{N_{\mathrm{X}}}\left(\dfrac{X_i}{\sigma_{X_i}}\right)^2, \quad 
B_{\mathrm{BAO}} = \sum_{i=1}^{N_{\mathrm{X}}}\dfrac{X_i X^{(0)}(z_i)}{\sigma_{X_i}^{2}}, \quad 
S_{\mathrm{BAO}}=\sum_{i=1}^{N_{\mathrm{X}}}\left(\dfrac{X^{(0)}(z_i)}{\sigma_{X_i}}\right)^2,
$$
and
$$
X^{(0)}(z_i) = r_d H_{0} X(z_i).
$$
Thus, we can determine the best-fit value of $H_{0}$ for given values of the other parameters:
$$
\nu^{(\mathrm{opt})} = \frac{B_{\mathrm{BAO}}}{S_{\mathrm{BAO}}}.
$$

\subsection{Hubble parameter test} 

In our analysis, we use the dependence of the Hubble parameter on redshift. From observations of galaxies, one can obtain differences in age and redshift for pairs of passively evolving galaxies. Assuming that the considered galaxies formed at the same time, we conclude that $H(z)=-(1+z)^{-1}\Delta z/\Delta t$ \cite{Jimenez:2001gg}. Table~\ref{tab:OHD} summarizes the data for the Hubble parameter $H(z_i)$ at various redshifts $z_i$ from different sources.

However, the uncertainties in these measurements can be substantial, as they depend strongly on various assumptions regarding stellar population synthesis. Uncertainties in the initial mass function and stellar metallicity may contribute up to 20--30\% to the total error budget \cite{Montiel:2020rnd,Moresco:2022phi,Rom:2023kqm}.

The best-fit parameters for the $H(z)$ dataset correspond to the minimum of the $\chi^2_H$ criterion:
\begin{equation}
\label{loglikeOHD}
    \chi^2_{H} = \sum_{i=1}^{36}\left[\dfrac{H_i-H(z_i)}{\sigma_{H_i}}\right]^2.
\end{equation}
As in the previous cases, we can determine the value of $H_0$ for which $\chi^2_{H}$ attains its minimum. Rewriting the expression, we obtain the following relation for $\chi^2_{H}$:
$$
\chi^{2}_{H}=A_{H}-2B_{H}H_{0}+H_{0}^{2}S_{H},
$$
$$
A_{H}=\sum_{i}\frac{H_{i}^{2}}{\sigma^{2}_{H_i}},\quad 
B_{H}=\sum_{i}\frac{E(z_{i})H_{i}}{\sigma^{2}_{H_i}},\quad
$$
$$
S_{H}=\sum_{i}\frac{E^{2}(z_i)}{\sigma^{2}_{H_i}},
$$
from which it follows that the minimum value of $\chi^2_{H}$ is
$$
{\bar{\chi}}_{H}^{2}=A_{H}-B_{H}^{2}/S_{H}
$$
at $H_{0}^{(\mathrm{opt})}={B_{H}/S_{H}}$.

\begin{table}[ht!]
\centering
\setlength{\tabcolsep}{1.5em}
\renewcommand{\arraystretch}{1.1}
\begin{tabular}{|c|c|c|c|}
   \hline
   Index & $z$ & $H(z)$ (km\,s$^{-1}$\,Mpc$^{-1}$) & Ref. \\
    \hline
1 & 0.0708  & $69.0\pm 19.6\pm12.4^\star$ & \cite{Zhang:2012mp} \\
2 & 0.09    & $69.0 \pm12.0\pm11.4^\star$  & \cite{Jimenez:2001gg} \\
3 & 0.12    & $68.6\pm26.2\pm11.4^\star$  & \cite{Zhang:2012mp} \\
4 & 0.17    & $83.0\pm8.0\pm13.1^\star$   & \cite{Simon:2004tf} \\
5 & 0.1791  & $75.0  \pm 3.8\pm0.5^\dagger$   & \cite{Moresco:2012jh} \\
6 & 0.1993  & $75.0\pm4.9\pm0.6^\dagger$   & \cite{Moresco:2012jh} \\
7 & 0.20    & $72.9\pm29.6\pm11.5^\star$  & \cite{Zhang:2012mp} \\
8 & 0.240   & $79.69\pm 2.65$ & \cite{Zhang:2012mp} \\
9 & 0.27    & $77.0\pm14.0\pm12.1^\star$  & \cite{Simon:2004tf} \\
10 & 0.28   & $88.8\pm36.6\pm13.2^\star$  & \cite{Zhang:2012mp} \\
11 & 0.3519 & $83.0\pm13.0\pm4.8^\dagger$  & \cite{Moresco:2016mzx} \\
12 & 0.3802 & $83.0\pm4.3\pm12.9^\dagger$  & \cite{Moresco:2016mzx} \\
13 & 0.4    & $95.0\pm17.0\pm12.7^\star$  & \cite{Simon:2004tf} \\
14 & 0.4004 & $77.0\pm2.1\pm10.0^\dagger$  & \cite{Moresco:2016mzx} \\
15 & 0.4247 & $87.1\pm2.4\pm11.0^\dagger$  & \cite{Moresco:2016mzx} \\
16 & 0.430  & $86.45\pm 3.68$    & \cite{Moresco:2016mzx} \\
17 & 0.4497 & $92.8 \pm4.5\pm 12.1^\dagger$  & \cite{Moresco:2016mzx} \\
18 & 0.47   & $89.0\pm23.0\pm44.0^\dagger$     & \cite{Ratsimbazafy:2017vga}\\
19 & 0.4783 & $80.9\pm2.1\pm 8.8^\dagger$   & \cite{Moresco:2016mzx} \\
20 & 0.48   & $97.0\pm62.0\pm12.7^\star$  & \cite{Stern:2009ep} \\
21 & 0.5929 & $104.0\pm11.6\pm4.5^\dagger$  & \cite{Moresco:2012jh} \\
22 & 0.6797 & $92.0\pm6.4\pm4.3^\dagger$   & \cite{Moresco:2012jh} \\
23 & 0.75   & $98.8\pm24.8\pm22.7^\dagger$     & \cite{Borghi:2021rft}\\
24 & 0.7812 & $105.0\pm9.4\pm6.1^\dagger$  & \cite{Moresco:2012jh} \\
25 & 0.80   & $113.1\pm15.1\pm20.2^\star$    & \cite{Jiao:2022aep}\\
26 & 0.8754 & $125.0\pm15.3\pm6.0^\dagger$  & \cite{Moresco:2012jh} \\
27 & 0.88   & $90.0\pm40.0\pm10.1^\star$  & \cite{Stern:2009ep} \\
28 & 0.9    & $117.0\pm23.0\pm13.1^\star$  & \cite{Simon:2004tf} \\
29 & 1.037  & $154.0\pm13.6\pm14.9^\dagger$  & \cite{Moresco:2012jh} \\
30 & 1.26   & $135.0\pm60.0\pm27.0^\dagger$   & \cite{Tomasetti:2023kek} \\
31 & 1.3    & $168.0\pm17.0\pm14.0^\star$  & \cite{Simon:2004tf} \\
32 & 1.363  & $160.0 \pm 33.6^\ddag$  & \cite{Moresco:2015cya} \\
33 & 1.43   & $177.0\pm18.0\pm14.8^\star$  & \cite{Simon:2004tf} \\
34 & 1.53   & $140.0\pm14.0\pm11.7^\star$  & \cite{Simon:2004tf} \\
35 & 1.75   & $202.0\pm40.0\pm16.9^\star$  & \cite{Simon:2004tf} \\
36 & 1.965  & $186.5 \pm 50.4^\ddag$  & \cite{Moresco:2015cya} \\
\hline
\end{tabular}
\caption{Redshifts, Hubble parameter measurements with uncertainties, and corresponding references. Systematic uncertainties are labeled with $\star$ when computed in this work, with $\dagger$ when taken from the literature, and with $\ddag$ when combined with statistical uncertainties.}
\label{tab:OHD}
\end{table}


\begin{table}[ht!]
\centering
\footnotesize
\setlength{\tabcolsep}{3pt}
\renewcommand{\arraystretch}{1.3}
\resizebox{\textwidth}{!}{%
\begin{tabular}{ll|ccc|ccc|ccc}
\hline
\textbf{Model$\backslash$Dataset} & &
\multicolumn{3}{c|}{\textbf{SNe Ia}} &
\multicolumn{3}{c|}{\textbf{$H(z)$}} &
\multicolumn{3}{c}{\textbf{BAO}} \\
\hline
$\Lambda$CDM & $\chi^2$ & \multicolumn{3}{c|}{545.093} & \multicolumn{3}{c|}{41.641} & \multicolumn{3}{c}{15.329} \\
 & $H_0$ & \multicolumn{3}{c|}{$72.55^{+0.68}_{-0.69}$} & \multicolumn{3}{c|}{$66.72^{+3.85}_{-4.10}$} & \multicolumn{3}{c}{$72.60^{+1.85}_{-8.47}$} \\
 & $\Omega_{de}$ & \multicolumn{3}{c|}{$0.634^{+0.052}_{-0.056}$} & \multicolumn{3}{c|}{$0.693^{+0.082}_{-0.105}$} & \multicolumn{3}{c}{$0.704^{+0.029}_{-0.032}$} \\
 & $r_d$ & \multicolumn{3}{c|}{$-$} & \multicolumn{3}{c|}{$-$} & \multicolumn{3}{c}{$140.31^{+14.69}_{-0.31}$} \\
\hline
$\Lambda$CDM, varying $G$ & $\chi^2$ & \multicolumn{3}{c|}{544.837} & \multicolumn{3}{c|}{41.443} & \multicolumn{3}{c}{14.930} \\
 & $H_0$ & \multicolumn{3}{c|}{$72.55^{+0.78}_{-0.82}$} & \multicolumn{3}{c|}{$66.62^{+4.45}_{-4.60}$} & \multicolumn{3}{c}{$66.64^{+7.97}_{-2.60}$} \\
 & $\Omega_{de}$ & \multicolumn{3}{c|}{$0.625^{+0.071}_{-0.078}$} & \multicolumn{3}{c|}{$0.791^{+0.093}_{-0.306}$} & \multicolumn{3}{c}{$0.814^{+0.028}_{-0.246}$} \\
 & $\alpha_G$ & \multicolumn{3}{c|}{$-0.098^{+0.198}_{-0.002}$} & \multicolumn{3}{c|}{$-0.099^{+0.199}_{-0.001}$} & \multicolumn{3}{c}{$-0.099^{+0.199}_{-0.001}$} \\
 & $r_d$ & \multicolumn{3}{c|}{$-$} & \multicolumn{3}{c|}{$-$} & \multicolumn{3}{c}{$153.44^{+1.56}_{-13.44}$} \\
\hline
\shortstack[l]{THDE, $G=\mbox{const}$} & $\chi^2$ & \multicolumn{3}{c|}{543.529} & \multicolumn{3}{c|}{37.819} & \multicolumn{3}{c}{15.435} \\
 & $H_0$ & \multicolumn{3}{c|}{$72.31^{+1.02}_{-0.83}$} & \multicolumn{3}{c|}{$77.74^{+12.25}_{-14.23}$} & \multicolumn{3}{c}{$67.38^{+14.53}_{-5.42}$} \\
 & $\Omega_{de}$ & \multicolumn{3}{c|}{$0.780^{+0.069}_{-0.225}$} & \multicolumn{3}{c|}{$0.749^{+0.099}_{-0.140}$} & \multicolumn{3}{c}{$0.707^{+0.070}_{-0.044}$} \\
 & $\gamma$ & \multicolumn{3}{c|}{$1.05^{+0.85}_{-0.35}$} & \multicolumn{3}{c|}{$1.70^{+0.20}_{-1.00}$} & \multicolumn{3}{c}{$1.89^{+0.01}_{-1.19}$} \\
 & $C$ & \multicolumn{3}{c|}{$1.49^{+0.01}_{-0.95}$} & \multicolumn{3}{c|}{$0.50^{+1.00}_{-0.00}$} & \multicolumn{3}{c}{$0.83^{+0.67}_{-0.33}$} \\
 & $r_d$ & \multicolumn{3}{c|}{$-$} & \multicolumn{3}{c|}{$-$} & \multicolumn{3}{c}{$151.55^{+3.45}_{-11.55}$} \\
\hline
 & & $C{=}0.7$ & $C{=}1$ & $C{=}1.2$ & $C{=}0.7$ & $C{=}1$ & $C{=}1.2$ & $C{=}0.7$ & $C{=}1$ & $C{=}1.2$ \\
\hline
\shortstack[l]{THDE, varying $G$\\$\gamma=0.9$} & $\chi^2$ & 546.411 & 544.239 & 543.612 & 39.186 & 41.629 & 42.891 & 16.447 & 15.571 & 16.230 \\
 & $H_0$ & $72.67^{+0.85}_{-0.82}$ & $72.47^{+0.80}_{-0.77}$ & $72.31^{+0.88}_{-0.72}$ & $70.74^{+5.04}_{-6.78}$ & $67.05^{+4.23}_{-5.40}$ & $65.67^{+3.77}_{-5.05}$ & $71.79^{+6.28}_{-7.51}$ & $69.72^{+4.83}_{-6.80}$ & $66.39^{+6.60}_{-4.03}$ \\
 & $\Omega_{de}$ & $0.615^{+0.061}_{-0.062}$ & $0.704^{+0.048}_{-0.087}$ & $0.742^{+0.051}_{-0.089}$ & $0.833^{+0.066}_{-0.273}$ & $0.838^{+0.061}_{-0.280}$ & $0.846^{+0.054}_{-0.286}$ & $0.717^{+0.149}_{-0.125}$ & $0.843^{+0.027}_{-0.228}$ & $0.847^{+0.029}_{-0.202}$ \\
 & $\alpha_G$ & $-0.099^{+0.199}_{-0.001}$ & $0.098^{+0.002}_{-0.198}$ & $0.100^{+0.000}_{-0.200}$ & $-0.100^{+0.200}_{-0.000}$ & $-0.100^{+0.200}_{-0.000}$ & $-0.099^{+0.199}_{-0.001}$ & $0.014^{+0.086}_{-0.114}$ & $-0.100^{+0.200}_{-0.000}$ & $-0.100^{+0.188}_{-0.000}$ \\
 & $r_d$ & $-$ & $-$ & $-$ & $-$ & $-$ & $-$ & $145.26^{+9.74}_{-5.26}$ & $146.20^{+8.80}_{-6.20}$ & $150.47^{+4.53}_{-10.47}$ \\
\hline
 & & $C{=}0.7$ & $C{=}1$ & $C{=}1.2$ & $C{=}0.7$ & $C{=}1$ & $C{=}1.2$ & $C{=}0.7$ & $C{=}1$ & $C{=}1.2$ \\
\hline
\shortstack[l]{THDE, varying $G$\\$\gamma=1$} & $\chi^2$ & 546.322 & 544.169 & 543.589 & 39.226 & 41.694 & 42.942 & 16.396 & 15.510 & 16.159 \\
 & $H_0$ & $72.67^{+0.83}_{-0.82}$ & $72.46^{+0.81}_{-0.78}$ & $72.32^{+0.86}_{-0.74}$ & $70.66^{+5.23}_{-6.76}$ & $67.09^{+4.20}_{-5.52}$ & $65.49^{+3.97}_{-4.91}$ & $68.49^{+9.62}_{-4.28}$ & $71.23^{+3.39}_{-8.30}$ & $68.79^{+4.14}_{-6.54}$ \\
 & $\Omega_{de}$ & $0.613^{+0.061}_{-0.063}$ & $0.701^{+0.048}_{-0.088}$ & $0.740^{+0.049}_{-0.094}$ & $0.829^{+0.070}_{-0.276}$ & $0.836^{+0.064}_{-0.279}$ & $0.836^{+0.064}_{-0.284}$ & $0.714^{+0.150}_{-0.126}$ & $0.838^{+0.029}_{-0.228}$ & $0.842^{+0.029}_{-0.205}$ \\
 & $\alpha_G$ & $-0.099^{+0.199}_{-0.001}$ & $0.100^{+0.000}_{-0.200}$ & $0.099^{+0.001}_{-0.199}$ & $-0.099^{+0.199}_{-0.001}$ & $-0.100^{+0.200}_{-0.000}$ & $-0.099^{+0.199}_{-0.001}$ & $0.014^{+0.085}_{-0.114}$ & $-0.100^{+0.200}_{-0.000}$ & $-0.099^{+0.188}_{-0.001}$ \\
 & $r_d$ & $-$ & $-$ & $-$ & $-$ & $-$ & $-$ & $152.22^{+2.78}_{-12.21}$ & $142.95^{+12.05}_{-2.95}$ & $145.32^{+9.68}_{-5.32}$ \\
\hline
 & & $C{=}0.7$ & $C{=}1$ & $C{=}1.2$ & $C{=}0.7$ & $C{=}1$ & $C{=}1.2$ & $C{=}0.7$ & $C{=}1$ & $C{=}1.2$ \\
\hline
\shortstack[l]{THDE, varying $G$\\$\gamma=1.1$} & $\chi^2$ & 546.251 & 544.117 & 543.564 & 39.241 & 41.738 & 42.959 & 16.348 & 15.456 & 16.021 \\
 & $H_0$ & $72.68^{+0.80}_{-0.87}$ & $72.42^{+0.84}_{-0.76}$ & $72.32^{+0.86}_{-0.75}$ & $70.46^{+5.28}_{-6.57}$ & $66.58^{+4.73}_{-5.13}$ & $65.53^{+3.96}_{-5.00}$ & $70.96^{+7.10}_{-6.57}$ & $68.70^{+5.49}_{-5.89}$ & $66.82^{+6.11}_{-4.67}$ \\
 & $\Omega_{de}$ & $0.612^{+0.061}_{-0.065}$ & $0.696^{+0.050}_{-0.084}$ & $0.736^{+0.050}_{-0.095}$ & $0.827^{+0.071}_{-0.270}$ & $0.824^{+0.076}_{-0.277}$ & $0.833^{+0.067}_{-0.285}$ & $0.717^{+0.144}_{-0.130}$ & $0.835^{+0.029}_{-0.228}$ & $0.837^{+0.029}_{-0.206}$ \\
 & $\alpha_G$ & $-0.098^{+0.198}_{-0.002}$ & $0.100^{+0.000}_{-0.200}$ & $0.099^{+0.001}_{-0.199}$ & $-0.100^{+0.200}_{-0.000}$ & $-0.100^{+0.200}_{-0.000}$ & $-0.100^{+0.200}_{-0.000}$ & $0.010^{+0.090}_{-0.110}$ & $-0.099^{+0.199}_{-0.001}$ & $-0.100^{+0.189}_{-0.000}$ \\
 & $r_d$ & $-$ & $-$ & $-$ & $-$ & $-$ & $-$ & $147.06^{+7.94}_{-7.06}$ & $148.36^{+6.64}_{-8.35}$ & $149.45^{+5.55}_{-9.45}$ \\
\hline
 & & $C{=}0.7$ & $C{=}1$ & $C{=}1.2$ & $C{=}0.7$ & $C{=}1$ & $C{=}1.2$ & $C{=}0.7$ & $C{=}1$ & $C{=}1.2$ \\
\hline
\shortstack[l]{THDE, varying $G$\\$\gamma=1.2$} & $\chi^2$ & 546.183 & 544.066 & 543.555 & 39.278 & 41.761 & 42.961 & 16.293 & 15.370 & 15.887 \\
 & $H_0$ & $72.66^{+0.83}_{-0.86}$ & $72.43^{+0.82}_{-0.75}$ & $72.30^{+0.87}_{-0.72}$ & $70.48^{+5.26}_{-6.69}$ & $66.84^{+4.35}_{-5.40}$ & $65.56^{+3.81}_{-5.04}$ & $73.82^{+4.20}_{-9.67}$ & $70.82^{+3.66}_{-8.13}$ & $66.37^{+6.52}_{-4.13}$ \\
 & $\Omega_{de}$ & $0.607^{+0.064}_{-0.059}$ & $0.696^{+0.048}_{-0.089}$ & $0.732^{+0.049}_{-0.095}$ & $0.825^{+0.071}_{-0.277}$ & $0.826^{+0.073}_{-0.285}$ & $0.832^{+0.067}_{-0.288}$ & $0.712^{+0.149}_{-0.128}$ & $0.830^{+0.032}_{-0.229}$ & $0.833^{+0.028}_{-0.212}$ \\
 & $\alpha_G$ & $-0.099^{+0.199}_{-0.001}$ & $0.099^{+0.001}_{-0.199}$ & $0.099^{+0.001}_{-0.199}$ & $-0.099^{+0.199}_{-0.001}$ & $-0.100^{+0.200}_{-0.000}$ & $-0.099^{+0.199}_{-0.001}$ & $0.013^{+0.087}_{-0.113}$ & $-0.099^{+0.199}_{-0.001}$ & $-0.100^{+0.191}_{-0.000}$ \\
 & $r_d$ & $-$ & $-$ & $-$ & $-$ & $-$ & $-$ & $141.18^{+13.82}_{-1.18}$ & $143.68^{+11.32}_{-3.68}$ & $150.50^{+4.50}_{-10.50}$ \\
\hline
\shortstack[l]{THDE, varying $G$\\($C$ and $\gamma$ are not fixed)} & $\chi^2$ & \multicolumn{3}{c|}{543.394} & \multicolumn{3}{c|}{37.324} & \multicolumn{3}{c}{14.876} \\
 & $H_0$ & \multicolumn{3}{c|}{$72.20^{+1.22}_{-0.83}$} & \multicolumn{3}{c|}{$77.35^{+10.87}_{-14.45}$} & \multicolumn{3}{c}{$70.47^{+11.20}_{-8.76}$} \\
 & $\Omega_{de}$ & \multicolumn{3}{c|}{$0.792^{+0.068}_{-0.276}$} & \multicolumn{3}{c|}{$0.824^{+0.076}_{-0.305}$} & \multicolumn{3}{c}{$0.808^{+0.077}_{-0.244}$} \\
 & $\alpha_G$ & \multicolumn{3}{c|}{$0.094^{+0.006}_{-0.194}$} & \multicolumn{3}{c|}{$-0.082^{+0.182}_{-0.018}$} & \multicolumn{3}{c}{$-0.096^{+0.196}_{-0.004}$} \\
 & $\gamma$ & \multicolumn{3}{c|}{$0.85^{+1.05}_{-0.15}$} & \multicolumn{3}{c|}{$1.58^{+0.32}_{-0.88}$} & \multicolumn{3}{c}{$1.86^{+0.04}_{-1.16}$} \\
 & $C$ & \multicolumn{3}{c|}{$1.47^{+0.03}_{-0.97}$} & \multicolumn{3}{c|}{$0.51^{+0.99}_{-0.01}$} & \multicolumn{3}{c}{$1.10^{+0.40}_{-0.60}$} \\
 & $r_d$ & \multicolumn{3}{c|}{$-$} & \multicolumn{3}{c|}{$-$} & \multicolumn{3}{c}{$143.83^{+11.17}_{-3.83}$} \\
\hline
\end{tabular}}
\caption{Separate fits for the probes (SNe Ia, $H(z)$, BAO) for various models of THDE model with and without varying $G$ in comparison with $\Lambda$CDM model: the results for $\chi^2_{\min}$
and best-fit parameters with asymmetric $1\sigma$ uncertainties
(upper/lower bounds of the
$\Delta\chi^2 < \Delta_{1\sigma}(k)$ region are given along the chain). The numbers of parameters are $k=3$ for simple $\Lambda$CDM;
$k=4$ for $\Lambda$CDM with varying $G$ and THDE with varying $G$ and fixed $C$ and $\gamma$; $k=5$ for THDE with $G=\mbox{const}$; $k=6$ for the THDE with varying $G$, $C$ and $\gamma$.
The parameter $r_d \in [140,155]$~Mpc is varied in all fits; for SNe Ia and $H(z)$
it is unobservable --- its intervals reproduce the prior).}
\label{Tab_3}
\end{table}

We restricted the allowed range of the parameter $\alpha_G$ to $-0.1\leq \alpha_G\leq 0.1$. { This is consistent with constraints derived from Hulse-Taylor binary pulsar \cite{Damour}, \cite{kogan}}. {The upper bound for $|\alpha_G|$ from this data is 0.07. Therefore there is no sense to consider the greater values althoug we consider 0.1 as limit for $\alpha_G$. From our investigation follows for cosmological models this range of $\alpha_G$ is clearly acceptable.} Therefore our main goal is to show that for this range of $\alpha$ cosmological model is inherently valid. As reference model we used in our analysis simple $\Lambda$CDM model. For illustration the $\Lambda$CDM model with varying $G$ is also considered.

\begin{figure}
\centering
\includegraphics[scale=0.38]{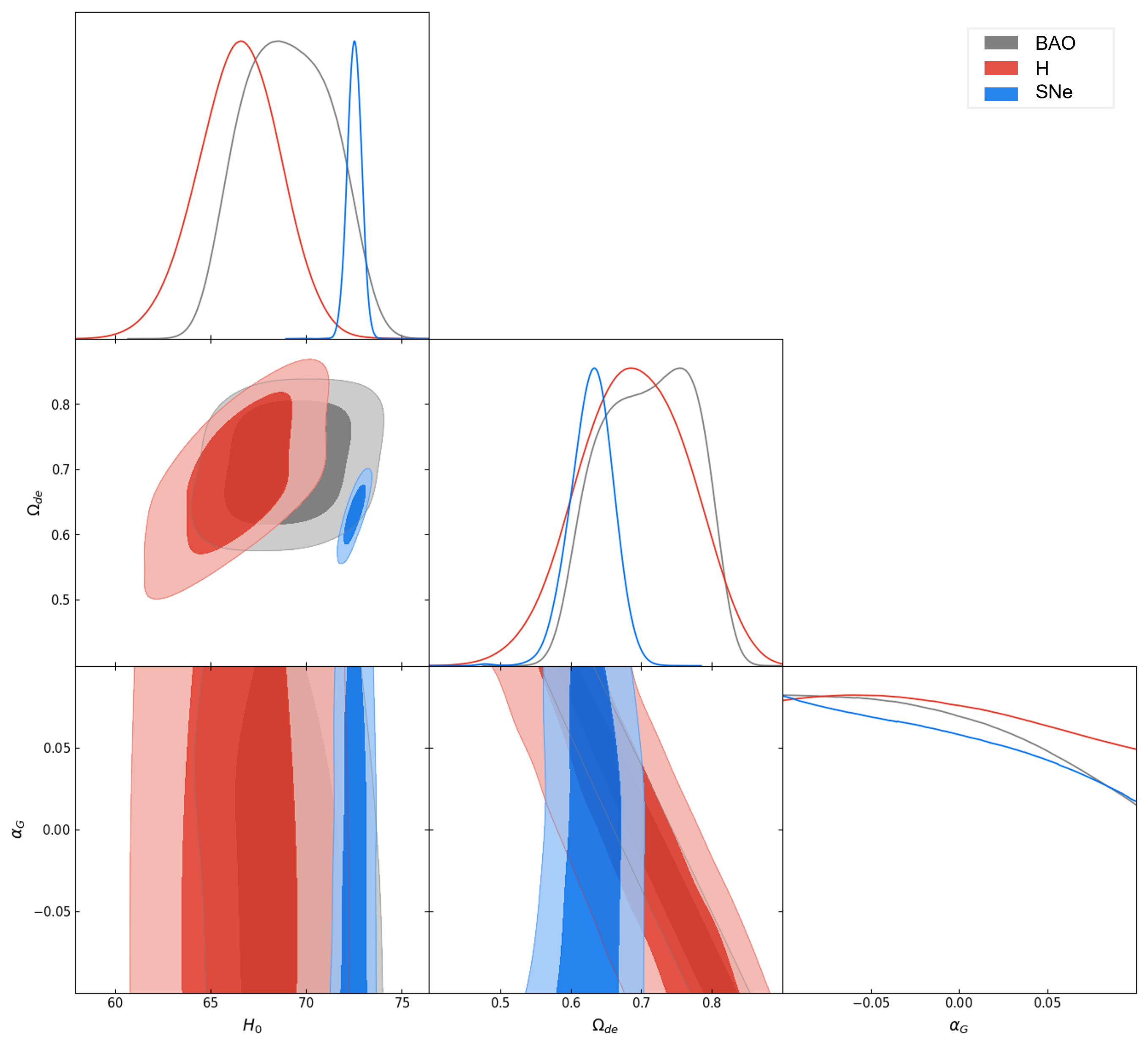}
\includegraphics[scale=0.38]{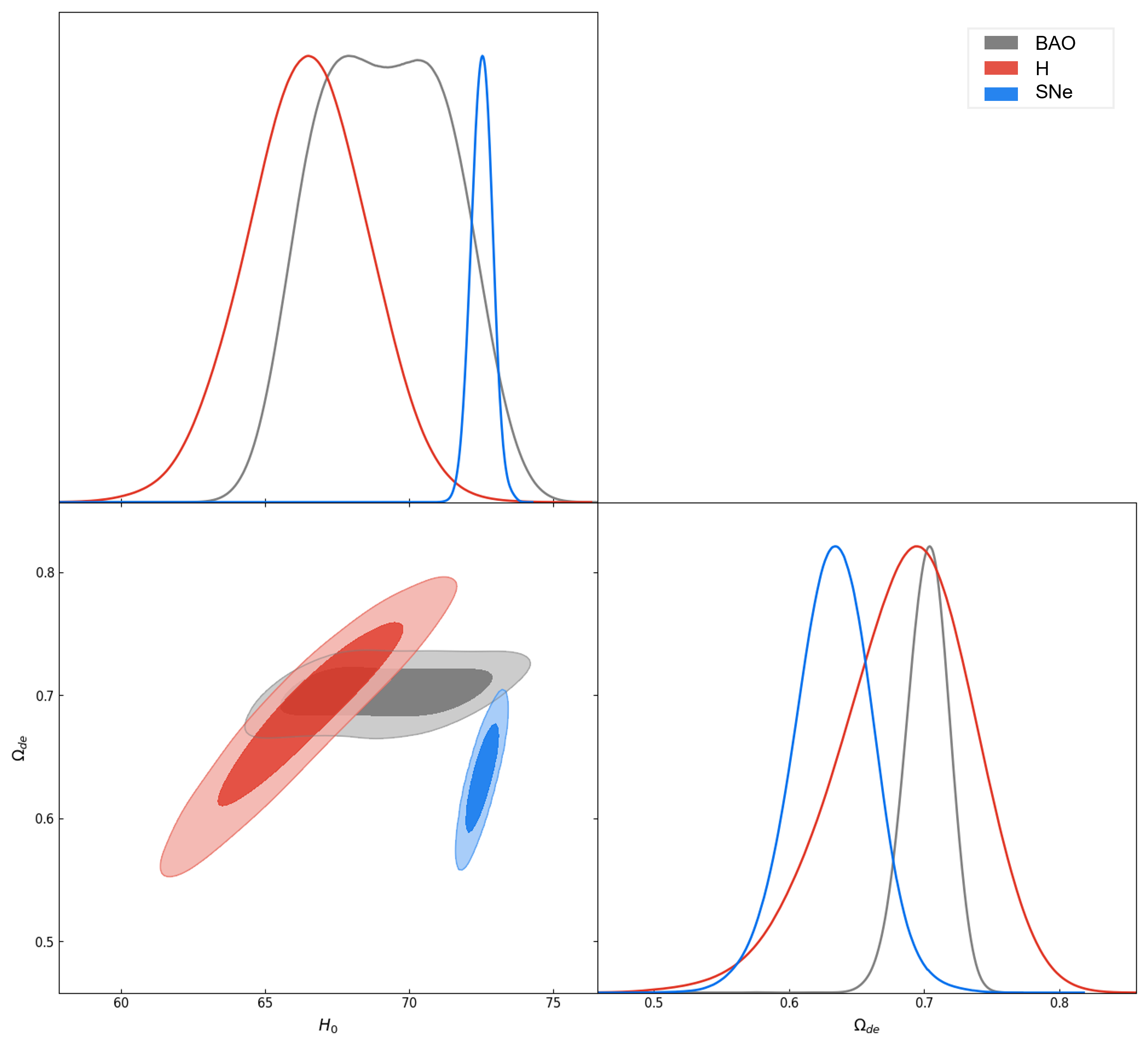}
\caption{The $1\sigma$ and $2\sigma$ confidence contours for parameters ($H_0$, $\Omega_{de}$, $r_d$, $\alpha_G$) from the three datasets in the case of the $\Lambda$CDM model with a varying gravitational constant (left panel). For comparison the results for simple $\Lambda$CDM model with three free parameters $H_0$, $\Omega_{de}$, $r_d$ are also given (right panel).}
\label{Fig5}
\end{figure}

For MCMC calculations we use a package Cobaya \cite{Cobaya}. From the data analysis (Table~\ref{Tab_3}), we observe the following notable features. Firstly one note that for some parameters allowed intervals coincide with prior boundaries therefore we have some asymmetry. The probes prefer opposite signs of $\alpha_G$ --- this is the
source of tension: $H(z)$ always pulls towards $-0.1$; SNe Ia towards $\alpha_G=0.1$
for THDE with varying $G$ with $C\ge1$ and $\Lambda$CDM, but towards to $-0.1$
for $C=0.7$; from the BAO data we have $\alpha_G = -0.1$ for $C\ge1$ and the internal optimum
$\alpha_G\approx0.01$ for $C=0.7$. 
From BAO data, $H_0$ is weakly constrained (this is due to the $r_d$--$H_0$ degeneracy): the large $\sigma$ and scatter of best-fit values along the valley are not convergence noise, but a feature of the probe. On Fig. \ref{Fig5} we depicted the confidence contours for separate analysis for three datasets for simple $\Lambda$CDM model and $\Lambda$CDM model with varying $G$. We clearly see Hubble tension between SNe Ia and $H(z)$ probes. For model with varying $G$ this tension is weaker.

\begin{figure}
\centering
\includegraphics[scale=0.38]{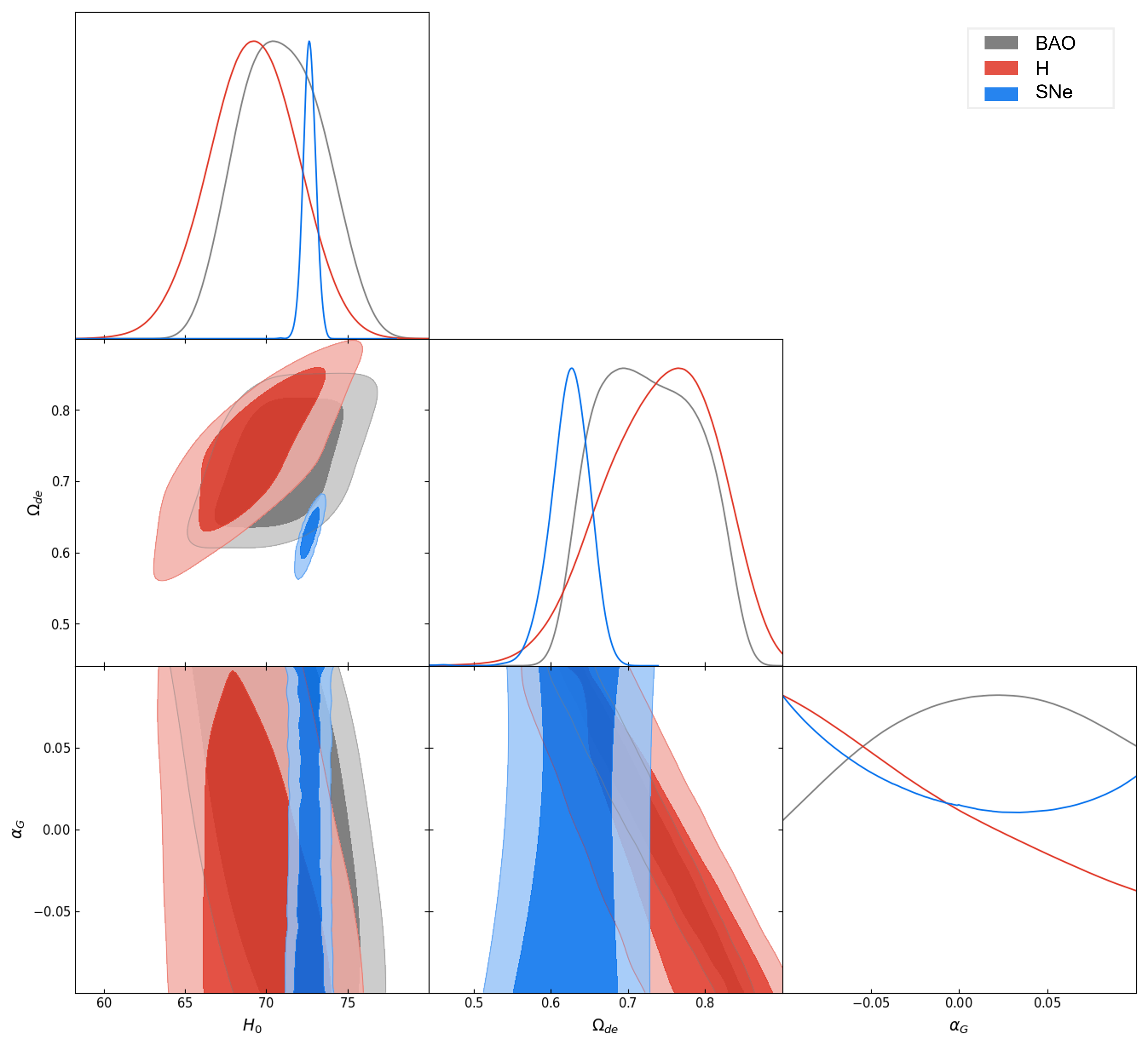}
\includegraphics[scale=0.38]{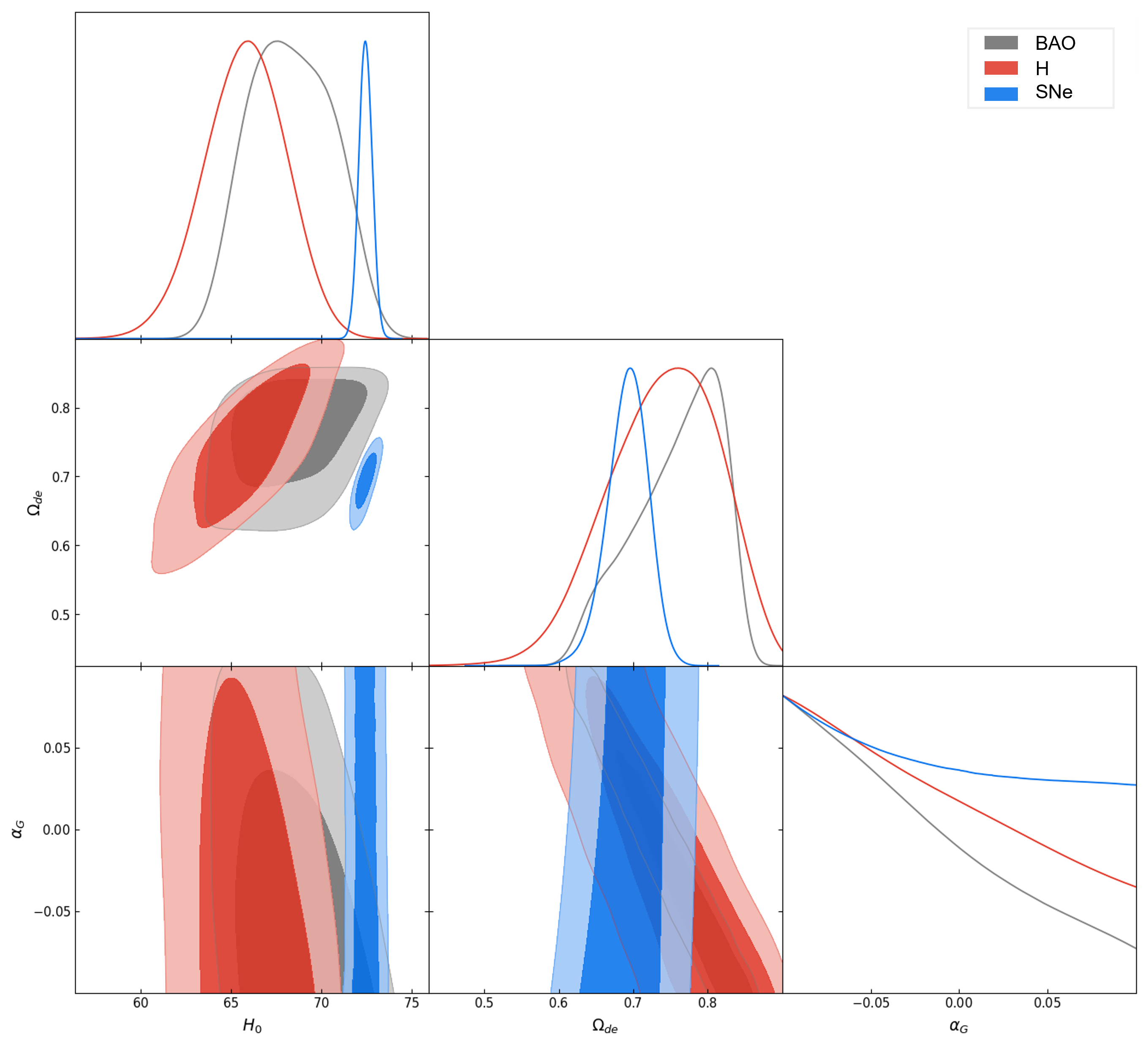}
\includegraphics[scale=0.38]{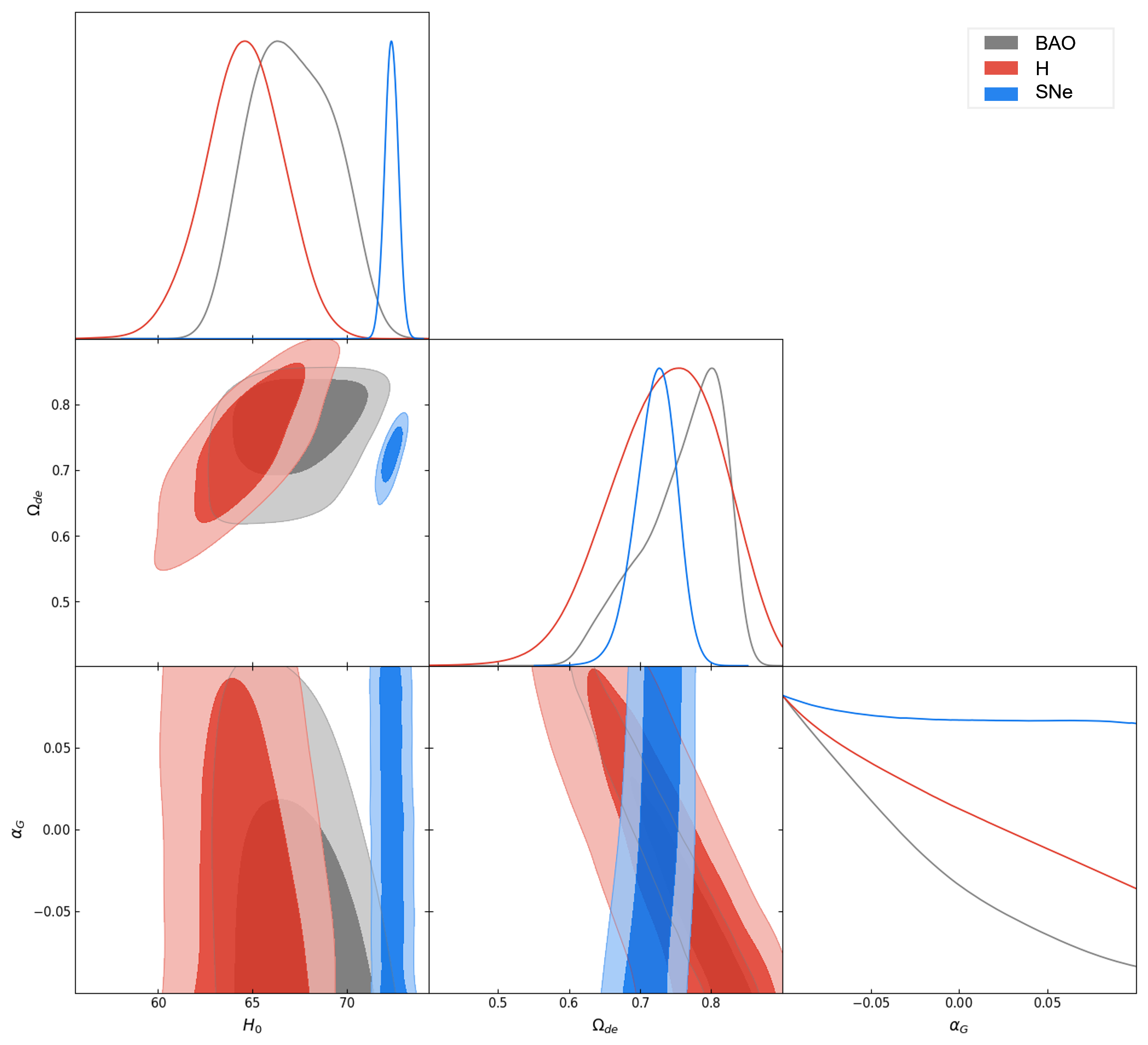}
\caption{The $1\sigma$ and $2\sigma$ confidence contours for parameters ($H_0$, $\Omega_{de}$, $r_d$, $\alpha_G$) from the three datasets in the case of the THDE model with a varying gravitational constant for varios $C$ and $\gamma$: $C=0.7$, $\gamma=0.9$ (left upper panel), $C=1$, $\gamma=1$ (right upper panel) and $C=1.2$, $\gamma=1.2$ (down panel).}
\label{Fig6}
\end{figure}

On Fig. \ref{Fig6} one can see the same for THDE model with some fixed $C$ and $\gamma$. The Hubble tension takes place in THDE models also. But there is a some tendency to its resolution for $C<1$. 

\begin{figure}
\centering
\includegraphics[scale=0.25]{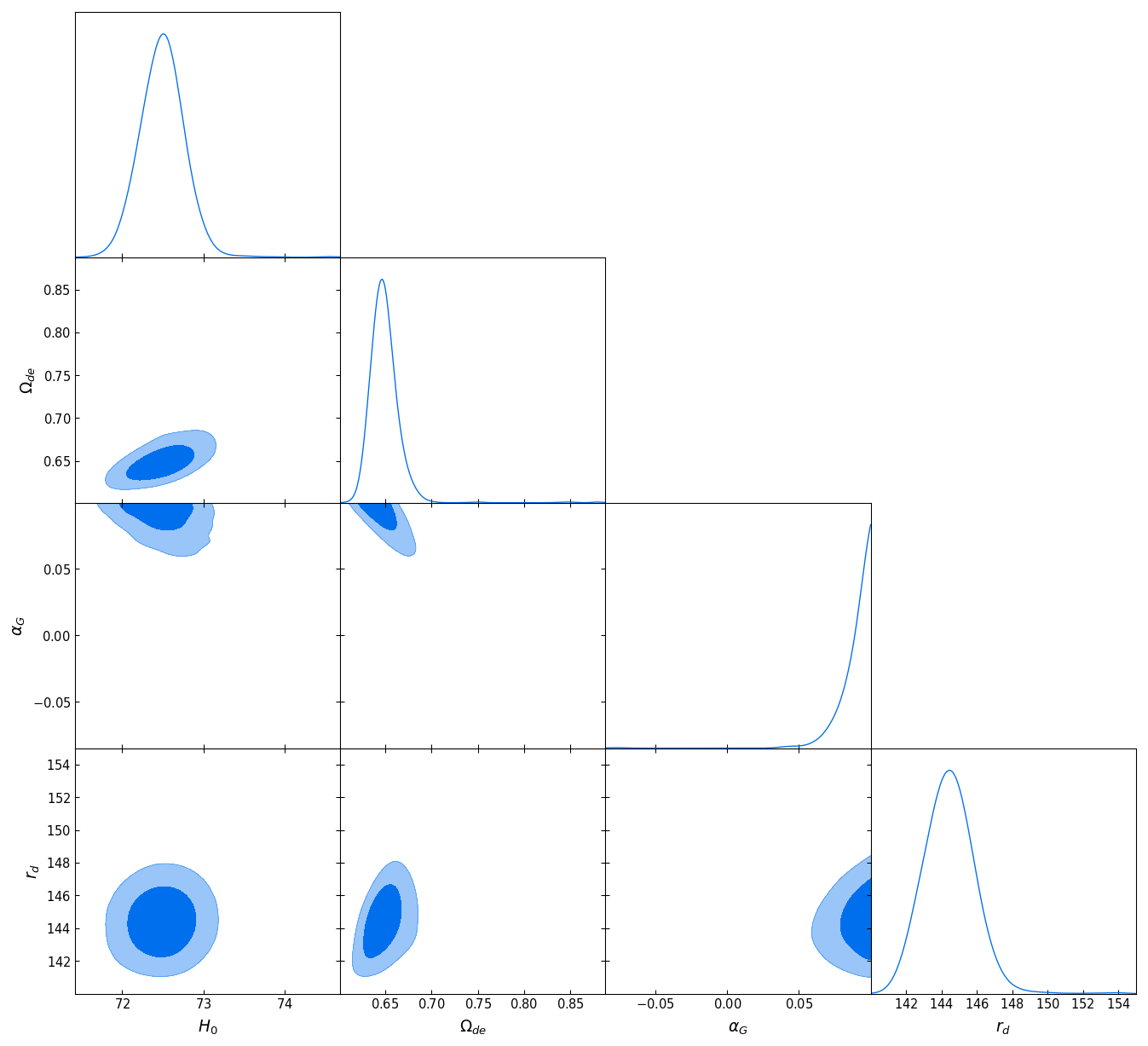}
\includegraphics[scale=0.25]{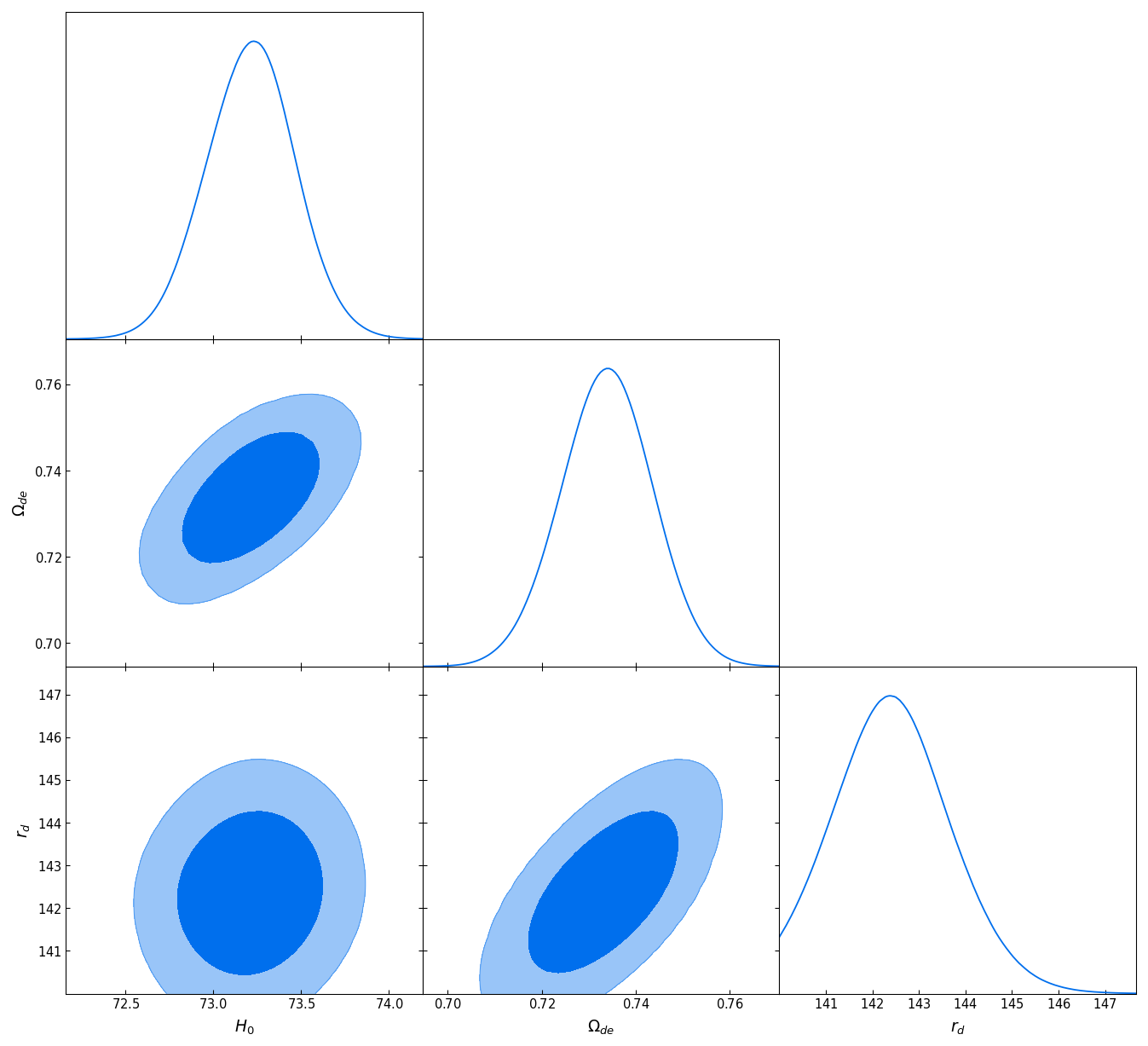}
\caption{The $1\sigma$ and $2\sigma$ confidence contours for parameters ($H_0$, $\Omega_{de}$, $r_d$, $\alpha_G$) from the combined analysis in the case of the $\Lambda$CDM model with a varying gravitational constant (upper panel). For comparison the results for simple $\Lambda$CDM model with three free parameters $H_0$, $\Omega_{de}$, $r_d$ are also given.}
\label{Fig7}
\end{figure}

And although this can be interpreted as that this model does not provide a consistent global cosmological fit the interesting feature appears for combined analysis of observational data.

\begin{figure}
\centering
\includegraphics[scale=0.25]{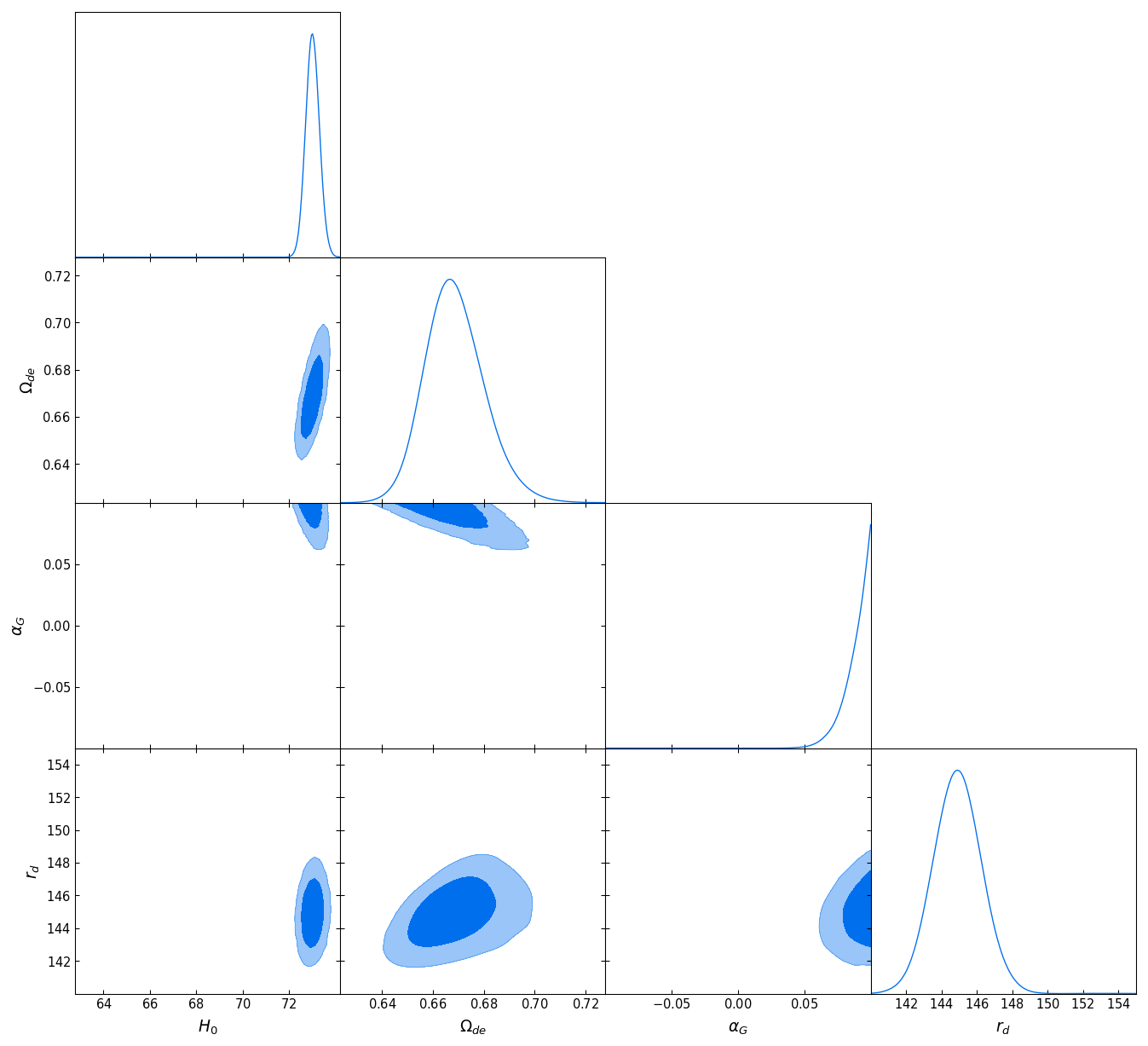}
\includegraphics[scale=0.25]{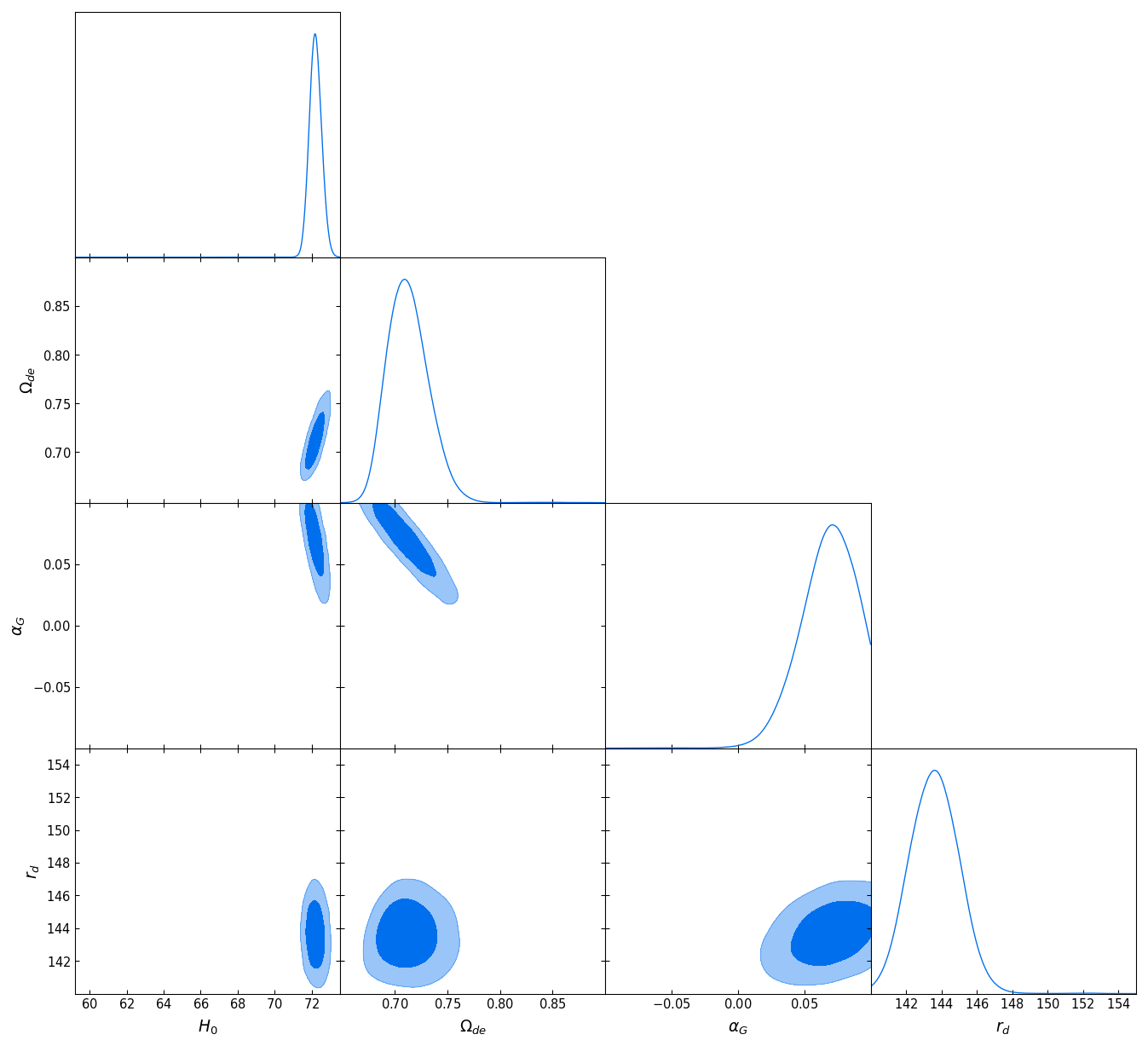}\\
\includegraphics[scale=0.25]{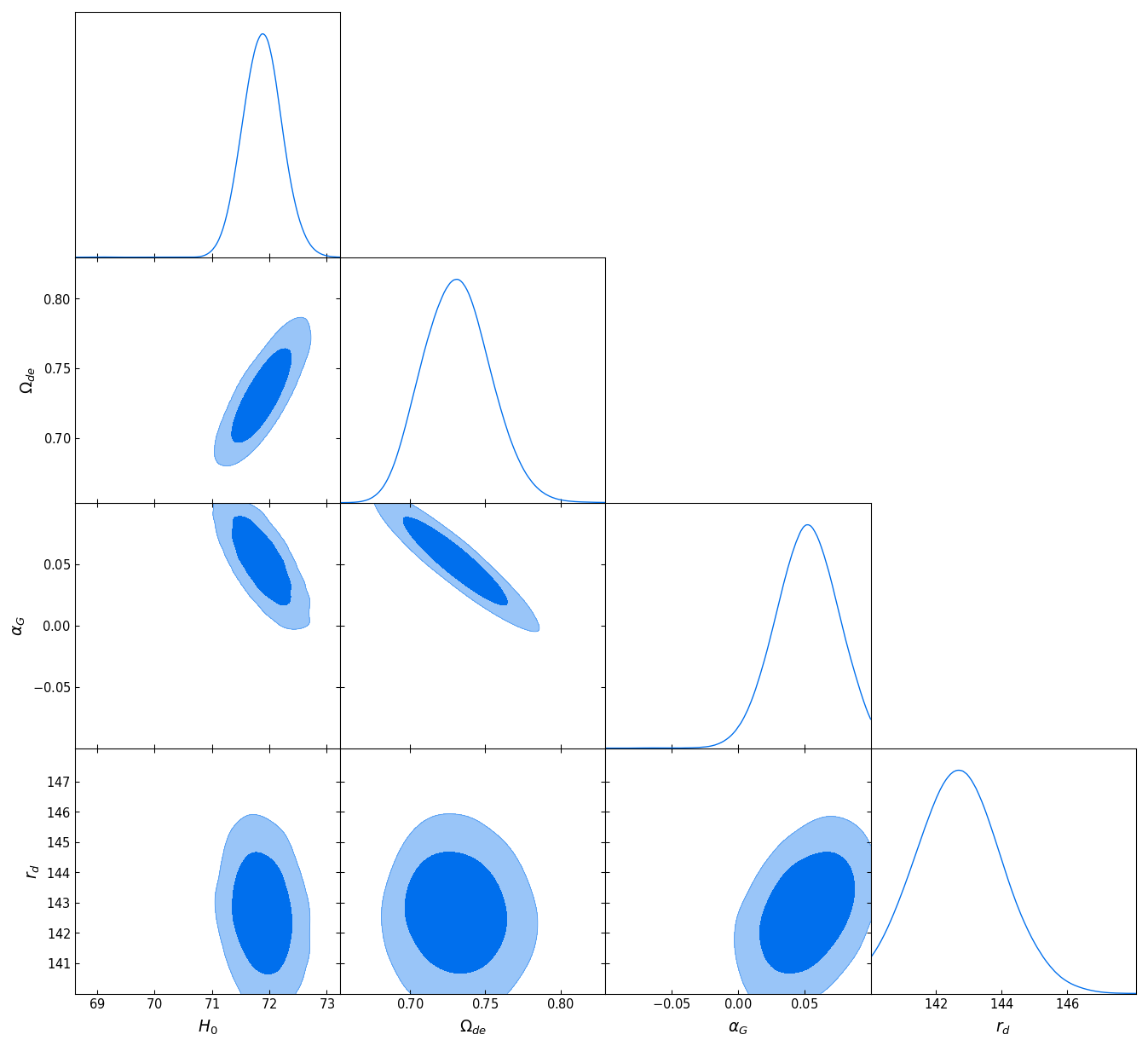}

\caption{The $1\sigma$ and $2\sigma$ confidence contours for parameters from the combined analysis in the case of the THDE model for some $C$ and $\gamma$: $C=0.7$, $\gamma=0.9$ (left upper panel), $C=1$, $\gamma=1$ (right upper panel), $C=1.2$, $\gamma=1.2$ (down panel).}
\label{Fig8}
\end{figure}

\begin{table}[ht]
\centering
\small
\setlength{\tabcolsep}{4pt}
\renewcommand{\arraystretch}{1.35}
\begin{tabular}{l|c|c|c|c}
\hline
\multicolumn{5}{c}{\textbf{Base models}} \\
\hline
\textbf{Parameter} & \shortstack{$\Lambda$CDM,\\ $G=\mbox{const}$} & \shortstack{$\Lambda$CDM,\\ $G\neq\mbox{const}$} &
 \shortstack{THDE, \\$G=\mbox{const}$} &
 \shortstack{THDE, \\$G\neq\mbox{const}$} \\
\hline
$\chi^2$      & 664.898 & 639.602 & 662.466 & 629.830 \\
$H_0$         & $73.21^{+0.47}_{-0.47}$ & $72.40^{+0.55}_{-0.50}$ & $73.76^{+0.87}_{-0.92}$ & $73.37^{+1.08}_{-0.91}$ \\
$\Omega_{de}$ & $0.734^{+0.018}_{-0.020}$ & $0.639^{+0.033}_{-0.020}$ & $0.730^{+0.048}_{-0.025}$ & $0.636^{+0.046}_{-0.028}$ \\
$\alpha_G$    & --- & $0.100^{+0.000}_{-0.029}$ & --- & $0.100^{+0.000}_{-0.026}$ \\
$r_d$         & $142.41^{+2.45}_{-2.40}$ & $144.63^{+3.08}_{-2.61}$ & $142.42^{+3.30}_{-2.42}$ & $145.13^{+3.39}_{-3.38}$ \\
$\gamma$      & --- & --- & $1.77^{+0.13}_{-1.07}$ & $1.37^{+0.53}_{-0.67}$ \\
$C$           & --- & --- & $0.752^{+0.154}_{-0.126}$ & $0.596^{+0.134}_{-0.096}$ \\
\hline
\multicolumn{5}{c}{\textbf{THDE with varying $G$, fixed $C$ and $\gamma$}} \\
\hline
\textbf{Parameter} & $\gamma=0.9$ & $\gamma=1$ &
$\gamma=1.1$ & $\gamma=1.2$ \\
\hline
\multicolumn{5}{c}{$C=0.7$} \\
\hline
$\chi^2$ & 634.468 & 634.769 & 634.943 & 635.053 \\
$H_0$ & $72.90^{+0.62}_{-0.56}$ & $72.88^{+0.62}_{-0.57}$ & $72.85^{+0.63}_{-0.58}$ & $72.88^{+0.60}_{-0.58}$ \\
$\Omega_{de}$ & $0.660^{+0.028}_{-0.019}$ & $0.658^{+0.028}_{-0.019}$ & $0.655^{+0.030}_{-0.018}$ & $0.654^{+0.029}_{-0.019}$ \\
$\alpha_G$ & $0.100^{+0.000}_{-0.029}$ & $0.100^{+0.000}_{-0.030}$ & $0.100^{+0.000}_{-0.030}$ & $0.100^{+0.000}_{-0.030}$ \\
$r_d$ & $145.19^{+2.87}_{-2.82}$ & $145.41^{+2.62}_{-3.03}$ & $145.09^{+2.86}_{-2.74}$ & $145.19^{+2.81}_{-2.81}$ \\
\hline
\multicolumn{5}{c}{$C=1$} \\
\hline
$\chi^2$ & 665.809 & 666.515 & 666.980 & 667.203 \\
$H_0$ & $72.18^{+0.76}_{-0.74}$ & $72.15^{+0.78}_{-0.75}$ & $72.11^{+0.80}_{-0.73}$ & $72.16^{+0.75}_{-0.78}$ \\
$\Omega_{de}$ & $0.712^{+0.048}_{-0.039}$ & $0.709^{+0.047}_{-0.039}$ & $0.704^{+0.049}_{-0.038}$ & $0.704^{+0.047}_{-0.043}$ \\
$\alpha_G$ & $0.072^{+0.028}_{-0.049}$ & $0.073^{+0.027}_{-0.052}$ & $0.074^{+0.026}_{-0.053}$ & $0.070^{+0.030}_{-0.050}$ \\
$r_d$ & $143.74^{+3.12}_{-2.95}$ & $143.82^{+2.95}_{-3.04}$ & $143.84^{+2.89}_{-3.07}$ & $143.81^{+2.94}_{-3.10}$ \\
\hline
\multicolumn{5}{c}{$C=1.2$} \\
\hline
$\chi^2$ & 688.168 & 688.638 & 688.654 & 688.225 \\
$H_0$ & $71.91^{+0.77}_{-0.75}$ & $71.90^{+0.77}_{-0.76}$ & $71.89^{+0.76}_{-0.76}$ & $71.88^{+0.78}_{-0.76}$ \\
$\Omega_{de}$ & $0.745^{+0.049}_{-0.050}$ & $0.740^{+0.050}_{-0.050}$ & $0.736^{+0.049}_{-0.052}$ & $0.730^{+0.051}_{-0.050}$ \\
$\alpha_G$ & $0.051^{+0.049}_{-0.051}$ & $0.053^{+0.047}_{-0.053}$ & $0.050^{+0.050}_{-0.051}$ & $0.052^{+0.048}_{-0.052}$ \\
$r_d$ & $142.63^{+2.97}_{-2.63}$ & $142.74^{+2.86}_{-2.74}$ & $142.57^{+2.97}_{-2.56}$ & $142.73^{+2.91}_{-2.73}$ \\
\hline
\end{tabular}
\caption{Optimal parameters and $\chi^2_{\min}$ from the combined fit
of SNe~Ia + $H(z)$ + BAO for THDE with a varying gravitational
constant ($C$ and $\gamma$ are fixed, while $H_0$,
$\Omega_{de}$, $\alpha_G \in [-0.1, 0.1]$ and $r_d \in [140, 155]$~Mpc
are varied) and for reference models: flat $\Lambda$CDM
with 3 parameters ($H_0$, $\Omega_{de}$, $r_d$), $\Lambda$CDM with varying $G$ (4 parameters), THDE with $G=\mbox{const}$ (5 parameters -- $H_0$,
$\Omega_{de}$, $\gamma$, $C$, $r_d$) and THDE with varying $G$ (6 parameters). Uncertainties are asymmetric $1\sigma$ intervals:
the boundaries of the $\Delta\chi^2 < \Delta_{1\sigma}(k)$ region along the MCMC chain
($\Delta_{1\sigma} = 3.53$ for $k=3$; $4.72$ for $k=4$; $5.89$ for
$k=5$; $7.04$ for $k=6$), with chain convergence $R-1 < 0.001$.}
\label{Tab_4}
\end{table}

This analysis of the observational data is presented in Table~\ref{Tab_4}. For THDE with $C=0.7$ and
$\Lambda$CDM model with varying $G$, and the six-parameter THDE, the optimum of $\alpha_G$
lies at the upper boundary of the prior interval ($\alpha_G = 0.1$):
the value $+0.000$ is a boundary artifact, not a measured uncertainty.
In the six-parameter THDE model ($C \in [0.5, 1.5]$, $\gamma \in [0.7, 1.9]$),
the optimum of $C$ is pushed to the lower prior boundary ($C \to 0.5$), while $\gamma$
is effectively unconstrained by the data (the $1\sigma$ range covers almost
the entire prior) --- consistent with the weak $\gamma$-dependence of $\chi^2$
from the grid scan. Optimal parameters for the THDE models from the combined analysis of SNe Ia+BAO+$H(z)$ data with corresponding $1\sigma$ and $2\sigma$ uncertainties for $H_0$ and $\Omega_{de}$. For comparison, in the $\Lambda$CDM model the optimal values are $H_0 = 72.30$~km\,s$^{-1}$\,Mpc$^{-1}$, $\Omega_{\mathrm{de}} = 0.688$, $\alpha_{G} = 0.094$, with a corresponding $\chi^2 = 653.9$.

\begin{figure}
\centering
\includegraphics[scale=0.4]{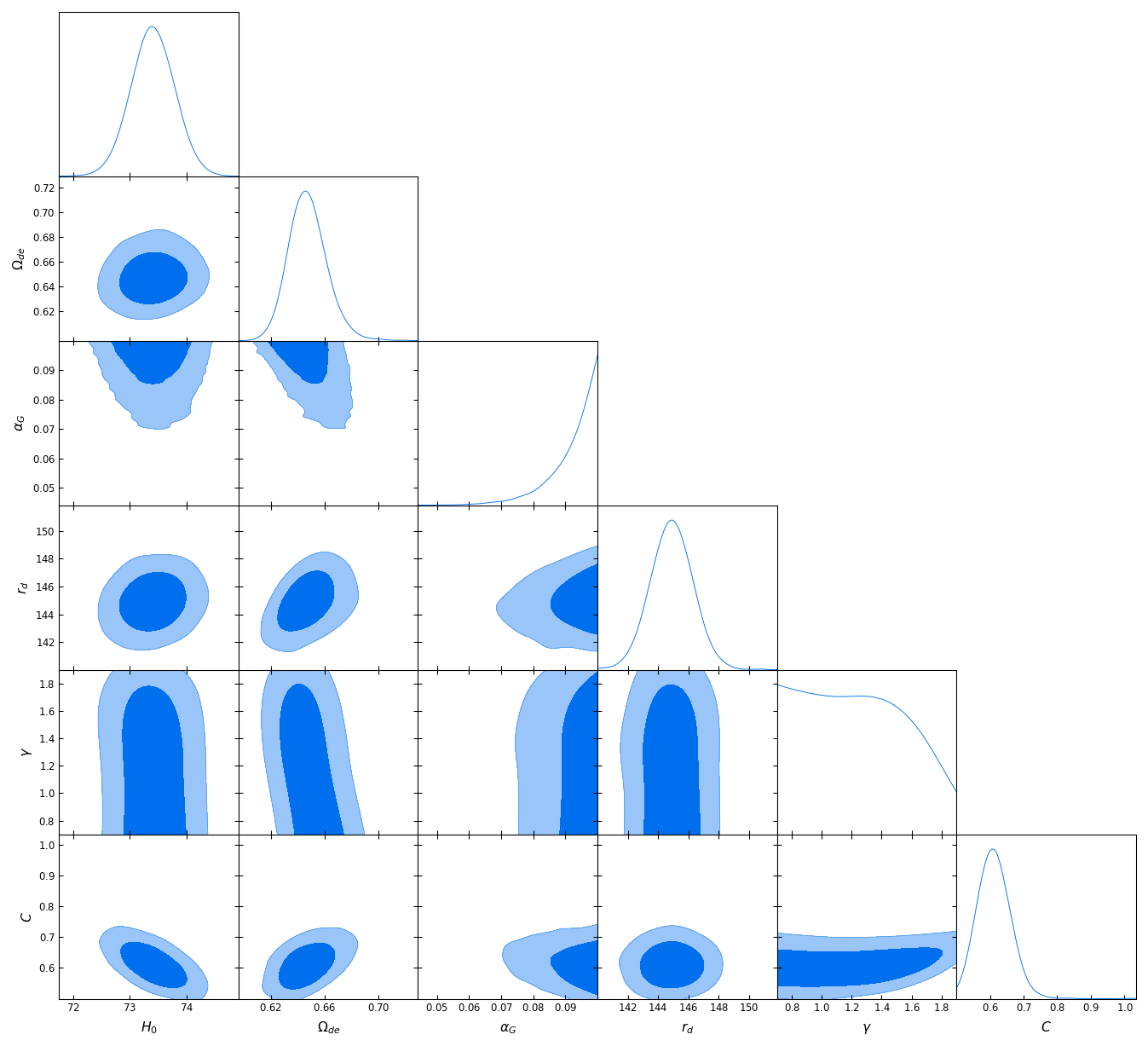}

\caption{The $1\sigma$ and $2\sigma$ confidence contours for parameters ($\Omega_{de}$, $r_d$, $H_0$, $\alpha_G$, $C$, $\gamma$) from the combined analysis in the case of the THDE model.}
\label{Fig9}
\end{figure}

The confidence contours for the parameters {are shown on Fig.~\ref{Fig7} and Fig. \ref{Fig8} for $\Lambda$CDM model and THDE model correspondingly. 

The best description of the data corresponds to THDE model with varying $G$ with $C=0.7$:
$\chi^2 \approx 634.5$, which is $\sim$5.1 better than $\Lambda$CDM model with varying $G$ (639.6) and $\sim$30.4 better than simple $\Lambda$CDM model (664.9) with the same number
of parameters as $\Lambda$CDM model with $G\neq \mbox{const}$ (4). The dependence from
$\gamma$ is weak ($\Delta\chi^2 \lesssim 0.6$ within a fixed $C$ block).

The location of the $\alpha_G$ optimum depends on $C$: for $C=0.7$ ---
at the prior boundary ($0.1$), for $C=1$ --- an internal minimum
$\alpha_G \approx 0.07$, for $C=1.2$ --- $\alpha_G \approx 0.05$.
For $C=0.7$, the true minimum lies beyond the prior, so
the $\chi^2$ reported in the table is an upper bound (cf.\ profile analysis:
extending the prior shifts the minimum towards $\alpha_G \approx 0.15$).

The $r_d$ values in all models are below the Planck
$147.09$~Mpc --- a manifestation of the $r_d$--$H_0$ degeneracy for
SH0ES-calibrated SNe Ia (high $H_0$ requires a lower $r_d$ from BAO). The $\chi^2_{\min}$ values are taken from the chains; a control refit
using the Nelder--Mead optimizer changes them by $\lesssim 0.05$.

\section{Comparison of models: tensions and inconsistency}

A single value of $\chi^2_{\min}$ is not sufficient for a full comparison
of the models; therefore, in this section we use several Bayesian-analysis
metrics~\cite{trotta2008}. The first of these is the Bayesian complexity:
\begin{equation}
p_D = \bar{D} - D(\hat\theta), \qquad D(\theta) = -2\ln L(\theta),
\label{eq:pD}
\end{equation}
where $D(\theta)$ is the deviance, $\bar{D}$ is its mean over the posterior
distribution, and $D(\hat\theta) = \chi^2_{\min}$ is the deviance at the
best-fit point. The complexity $p_D$ measures the effective number of
parameters that the data are able to constrain: for $p_D \approx k$
($k$ being the number of varied parameters), the data constrain all
parameters; $p_D < k$ means that some of the parameters are redundant
(not constrained by the data); $p_D > k$ is a sign of excessive fitting
flexibility (a tendency to adapt to noise; a small excess can also arise
for non-Gaussian posteriors truncated by prior boundaries). Instead of the
commonly used AIC and BIC criteria, which are approximate and do not always
adequately characterize a model~\cite{trotta2008}, we use the deviance
information criterion
\begin{equation}
\mathrm{DIC} = \bar{D} + p_D = D(\hat\theta) + 2p_D,
\label{eq:DIC}
\end{equation}
Unlike AIC/BIC, this criterion penalizes a model not for the nominal number
of parameters but for the effective number of parameters. We also use the
Bayesian evidence
\begin{equation}
Z = \int L(\theta)\,\pi(\theta)\,d\theta,
\label{eq:evidence}
\end{equation}
i.e., the likelihood averaged over the prior distribution $\pi(\theta)$:
the evidence automatically includes an ``Occam penalty''---a model with
unnecessary prior volume not supported by the data receives a smaller $Z$.
Models are compared using the Bayes factor $B_{i0} = Z_i/Z_0$; we report
its logarithm $\ln B_{i0} = \ln Z_i - \ln Z_{\Lambda\mathrm{CDM}}$
relative to the standard model. To compute the evidence, we used the
nested-sampling algorithm PolyChord~\cite{polychord2015}, which is
integrated with Cobaya.

Table~\ref{tab:bayes-comparison} presents the computed metrics for the standard cosmological model $\Lambda$CDM; $\Lambda$CDM with a varying
gravitational constant; THDE; THDE with a varying
gravitational constant and fixed parameters $C$ and $\gamma$; and the full
THDE model with all parameters varied; the values
of $\Delta\mathrm{DIC}$ and $\ln B_{i0}$ are given relative to the standard
model.

\begin{table}[ht]
\centering
\renewcommand{\arraystretch}{1.35}
\begin{tabular}{l|c|c|c|c|c}
\hline
\textbf{Model} & $\chi^2_{\min}$ & $k$ & $p_D$ &
$\Delta\mathrm{DIC}$ & $\ln B_{i0}$ \\
\hline
$\Lambda$CDM                        & 664.898 & 3 & 2.91 & 0        & 0 \\
$\Lambda$CDM$,\alpha_G$             & 639.602 & 4 & 4.64 & $-21.84$ & $+10.01$ \\
THDE              & 662.466 & 5 & 4.40 & $+0.55$  & $-1.09$ \\
THDE$,\alpha_G$, $C=0.7$, $\gamma=0.9$         & 634.468 & 4 & 4.57 & $-27.12$ & $+12.33$ \\
THDE$,\alpha_G$, $C=1$, $\gamma=1$             & 666.515 & 4 & 3.70 & $+3.20$  & $-2.36$ \\
THDE$,\alpha_G$, $C=1.2$, $\gamma=1.2$         & 688.225 & 4 & 3.74 & $+24.98$ & $-13.06$ \\
THDE$,\alpha_G$   & 629.830 & 6 & 5.83 & $-29.23$ & $+12.01$ \\
\hline
\end{tabular}
\caption{Bayesian comparison of the models for the joint fit
SNe~Ia + $H(z)$ + BAO.}
\label{tab:bayes-comparison}
\end{table}

For $\Delta\mathrm{DIC}$, negative values indicate that the model is
preferred over $\Lambda$CDM (thresholds: $\sim$2 --- weak, 5--7 --- moderate,
$>10$ --- strong preference). The Bayes factor is interpreted using the
Jeffreys scale in the calibration of~\cite{trotta2008}
(Table~\ref{tab:jeffreys}). An analysis of these metrics shows that the data
favour models with a varying gravitational constant: for all
leading models, $\Delta\mathrm{DIC}$ is approximately $-22$ or lower and
$\ln B_{i0} \ge +10$, which corresponds to a decisive level according to
both scales. The best metrics are obtained for the Tsallis holographic dark energy
model with $\alpha_G$ and $C<1$. As $C$ increases, the models provide a
noticeably worse description of the data and are disfavoured relative to the
standard cosmological model: for $C=1$, the fit quality is comparable to
that of $\Lambda$CDM ($\chi^2_{\min}=666.5$ versus $664.9$), although the
evidence is somewhat lower, whereas for $C=1.2$ the model is decisively
worse than $\Lambda$CDM according to both criteria. We note that fully
varying $C$ and $\gamma$ gives the best value of $\Delta\mathrm{DIC}$
($-29.23$) and almost the same logarithm of the Bayes factor
($+12.01$ versus $+12.33$ for fixed $C=0.7$): the likelihood gain almost
completely compensates the Occam penalty for the two additional parameters.
It is also worth emphasizing that the reported estimates are quite
conservative: the adopted variation range $\alpha_G \in [-0.1, 0.1]$ is
physically motivated by the pulsar constraint
$|\alpha_G| \lesssim 0.07$, and in all leading models the best-fit value of
$\alpha_G$ reaches the upper prior boundary, therefore, the reported
$\chi^2_{\min}$, DIC, and $\ln B$ should be regarded as upper estimates of
the preference for a varying $G$. Finally, comparing the $k$ and $p_D$
columns shows that the best parameter constraints are achieved for
$\Lambda$CDM ($p_D = 2.91$ for $k=3$) and for the full THDE
model ($p_D = 5.83$ for $k=6$), the data constrain practically all
parameters.

\begin{table}[ht]
\centering
\renewcommand{\arraystretch}{1.35}
\begin{tabular}{c|c|l}
\hline
$|\ln B_{i0}|$ & Odds & Interpretation \\
\hline
$<1$      & $\lesssim 3:1$   & inconclusive \\
$1$--$2.5$  & $\sim 3:1$     & weak preference \\
$2.5$--$5$  & $\sim 12:1$    & moderate preference \\
$>5$      & $\gtrsim 150:1$  & strong preference \\
\hline
\end{tabular}
\caption{Jeffreys scale for interpreting the Bayes factor
(in the calibration of~\cite{trotta2008}).}
\label{tab:jeffreys}
\end{table}

It is also of interest to examine how the models under consideration
address the tension problem in the data. For this analysis we use the
index of inconsistency (IOI), proposed and developed
in~\cite{lin_ishak_ioi,lin_ishak_bayes,garcia_quintero_ioi3}. We
analyze the discrepancy between the probes in the parameter subspace
$(H_0, \Omega_{de})$, which is common to all models; it is precisely
in this subspace that the Hubble tension is localized.
According to~\cite{lin_ishak_ioi}, the parameter $r_d$, which enters
only the BAO observables and is not constrained by SNe Ia and $H(z)$,
can be marginalized out without loss of information
(model-specific parameters $\alpha_G$, $\gamma$, $C$ are not included:
their posteriors for individual probes are almost unconstrained and
are truncated by the priors, and the IOI values would become
incomparable between models). Table~\ref{tab:tensions} gives the IOI
for each pair of probes, computed from
\begin{equation}
\mathrm{IOI} = \frac{1}{2}\,\delta^{T} (C_1 + C_2)^{-1} \delta,
\qquad \delta = \mu_1 - \mu_2,
\label{eq:ioi}
\end{equation}
where $\mu_i$ and $C_i$ are the means and covariance matrices of the
individual-probe posteriors in the chosen subspace. The table also
gives the multi-IOI for all probes simultaneously:
\begin{equation}
\mathrm{IOI} = \frac{1}{N}\sum_{i=1}^{N}
(\mu_i - \mu_F)^{T} F_i\, (\mu_i - \mu_F),
\qquad
F_F = \sum_{i=1}^{N} F_i,
\quad
\mu_F = F_F^{-1} \sum_{i=1}^{N} F_i \mu_i,
\label{eq:multi-ioi}
\end{equation}
where $F_i = C_i^{-1}$ are the Fisher matrices of the probes, and
$\mu_F$ is their weighted mean. For $N=2$, expression~(\ref{eq:multi-ioi})
reduces to the pairwise one~(\ref{eq:ioi}). To interpret the values of
this metric, the authors proposed a Jeffreys-like scale
(Table~\ref{tab:ioi-scale}). We note that varying $r_d$ removes the
``classical'' tension between the SNe and BAO probes; however, as seen
from the table, the tension between SNe Ia and $H(z)$ is present and is
dominant.

\begin{table}[ht]
\centering
\small
\setlength{\tabcolsep}{4pt}
\renewcommand{\arraystretch}{1.5}
\begin{tabular}{l|c|c|c|c|c|c|c|c|c|c}
\hline
\textbf{Model} &
\shortstack{IOI\\(SNe Ia, BAO)} &
\shortstack{IOI\\($H(z)$, BAO)} &
\shortstack{IOI\\(SNe Ia, $H(z)$)} &
\shortstack{IOI\\(multi)} &
$Q_{\mathrm{DMAP}}$ & $k_{\mathrm{eff}}$ & $\sim\!\sigma$ & $\ln S$ & $d$ & $\sim\!\sigma_S$ \\
\hline
$\Lambda$CDM            & 4.61 & 0.43 & 27.17 & 21.44 & 62.8 & 2.97 & 7.4 & $-29.85$ & 3.35 & 7.3 \\
$\Lambda$CDM$,\alpha_G$ & 2.35 & 0.38 &  9.13 &  7.40 & 38.4 & 1.98 & 5.9 & $-18.09$ & 1.61 & 5.9 \\
THDE  & 0.74 & 0.02 & 0.26 & 0.61 & 65.7 & 2.90 & 7.6 & $-29.91$ & 4.71 & 7.1 \\
\shortstack[l]{THDE$,\alpha_G$,\\$C=0.7$, $\gamma=0.9$} & 2.37 & 0.26 &  8.11 &  6.47 & 32.4 & 2.07 & 5.3 & $-14.78$ & 1.89 & 5.3 \\
\shortstack[l]{THDE$,\alpha_G$,\\$C=1$, $\gamma=1$}     & 3.29 & 0.35 & 13.27 & 10.56 & 65.1 & 2.28 & 7.7 & $-30.30$ & 3.15 & 7.4 \\
\shortstack[l]{THDE$,\alpha_G$,\\$C=1.2$, $\gamma=1.2$} & 3.84 & 0.42 & 16.48 & 13.15 & 85.8 & 2.39 & 8.9 & $-40.56$ & 3.52 & 8.6 \\
THDE$,\alpha_G$ & 0.86 & 0.02 &  0.38 &  0.75 & 34.2 & 1.89 & 5.5 & $-14.31$ & 3.24 & 4.9 \\
\hline
\end{tabular}
\caption{Tension metrics for the SNe~Ia, $H(z)$, and BAO probes.}
\label{tab:tensions}
\end{table}

The table shows a high level of tension for the $\Lambda$CDM model;
this tension weakens when a varying gravitational constant is taken
into account. The plain THDE model, like the full THDE$+\alpha_G$
model, would seem to reduce the tension between the probes, and quite
substantially. However, this reveals a known problem of the IOI
metric: as the number of model parameters increases, the
individual-probe posteriors broaden, and the metric is
underestimated~\cite{lin_ishak_ioi}; the IOI decreases both when the
means genuinely move closer together and when the uncertainties simply
increase. For a more detailed analysis of the tension, we consider an
additional metric proposed in~\cite{raveri_hu} and systematically used
for ranking cosmological models~\cite{leizerovich_review}, \cite{lemos_des}:
\begin{equation}
Q_{\mathrm{DMAP}} = \chi^2_{\min}(\mathrm{joint}) -
\sum_i \chi^2_{\min}(i)
\label{eq:qdmap}
\end{equation}
This metric quantifies the loss of fit quality when combining the
probes; it compares not the allowed regions of parameter space but the
best fits, and is therefore insensitive to the width of the
posteriors. The values are given in Table~\ref{tab:tensions}. To
convert them into Gaussian equivalents, we also computed the effective
number of degrees of freedom following~\cite{raveri_hu}:
\begin{equation}
N_{\mathrm{eff}} = N - \mathrm{tr}\!\left(C_\pi^{-1} C_p\right),
\qquad
k_{\mathrm{eff}} = \sum_i N_{\mathrm{eff},i} - N_{\mathrm{eff,joint}},
\label{eq:keff}
\end{equation}
where $N$ is the number of varied parameters of the model, and
$C_\pi$ and $C_p$ are the covariance matrices of the prior and
posterior (weakly constrained parameters do not contribute to
$N_{\mathrm{eff}}$); the significance $\sim\!\sigma$ is obtained under
the assumption $Q_{\mathrm{DMAP}} \sim \chi^2_{k_{\mathrm{eff}}}$.

It is immediately apparent that the weakening of the tension in the IOI
metric for the THDE model with $G=\mbox{const}$ is an artifact. The
$Q_{\mathrm{DMAP}}$ metric gives a tension estimate at the level of the
$\Lambda$CDM model ($65.7$ versus $62.8$, i.e. $7.6\sigma$ versus
$7.4\sigma$). At the same time, the reduction in tension indicated by
IOI when a varying gravitational constant is added is also confirmed by
the decrease in $Q_{\mathrm{DMAP}}$: $62.8 \to 38.4$ ($5.9\sigma$) for
$\Lambda$CDM with varying $G$ and down to $32.4$ ($5.3\sigma$) for
THDE with varying $G$ and $C=0.7$. Here we again see a preference for the
THDE model with a varying gravitational constant and $C<1$ (for
$C\ge1$, the tension instead increases up to $8.9\sigma$). However,
the tension is only partially resolved.

\begin{table}[htbp]
\centering
\renewcommand{\arraystretch}{1.35}
\begin{tabular}{c|l}
\hline
IOI & Interpretation \\
\hline
$<1$      & no significant inconsistency \\
$1$--$2.5$  & weak inconsistency \\
$2.5$--$5$  & moderate inconsistency \\
$>5$      & strong inconsistency \\
\hline
\end{tabular}
\caption{Jeffreys-like scale for interpreting the IOI
index~\cite{lin_ishak_ioi}.}
\label{tab:ioi-scale}
\end{table}

Another useful metric for cross-checking the results is
suspiciousness~\cite{handley_lemos_susp,handley_lemos_dim,
lemos_des}, along with the associated Bayesian dimensionality. This
metric is based on Bayesian evidence values obtained by nested sampling
(the calculations were performed with the anesthetic
package~\cite{anesthetic}, and the results are given in
Table~\ref{tab:tensions}). It takes into account the full prior
volume, rather than only the posterior means and covariances.
Suspiciousness is defined as
\begin{equation}
\ln S = \ln R - \ln I,
\label{eq:suspiciousness}
\end{equation}
where $\ln R = \ln Z_{\mathrm{joint}} - \sum_i \ln Z_i$ is the
logarithm of the Bayes ratio of the joint fit to the individual probes,
and $\ln I = \sum_i \mathcal{D}_i - \mathcal{D}_{\mathrm{joint}}$ is
its information component, expressed through the Kullback--Leibler
divergence $\mathcal{D} = \int \mathcal{P}\ln(\mathcal{P}/\pi)\,
d\theta$ between the posterior $\mathcal{P}$ and the prior $\pi$. By
itself, the ratio $R$ depends strongly on the arbitrary choice of prior
widths. Subtracting $\ln I$ removes this dependence and leaves a
genuine measure of data agreement: $\ln S \approx 0$ indicates that the
probes are consistent, while a significantly negative $\ln S$ indicates
tension. The Bayesian dimensionality is computed as
\begin{equation}
d = \sum_i d_{G,i} - d_{G,\mathrm{joint}},
\qquad
\frac{d_G}{2} = \left\langle \left(\ln L -
\langle \ln L \rangle_{\mathcal{P}} \right)^2 \right\rangle_{\mathcal{P}},
\label{eq:bmd}
\end{equation}
where $d_G$ is the Bayesian dimensionality of the
model~\cite{handley_lemos_dim} (twice the variance of $\ln L$ over the
posterior), i.e. the measured effective number of parameters
constrained by that data set. The quantity $d$ measures the number of
parameters constrained jointly by the probes, i.e. the number of
effective degrees of freedom in which the probes can conflict. The
significance of the tension is obtained from the statistic
$d - 2\ln S \sim \chi^2_d$.

In the Gaussian limit, the quantity $d - 2\ln S$ corresponds to the
$Q_{\mathrm{DMAP}}$ statistic. For all models, the Bayesian
inconsistency measure confirms the frequentist estimate. The quantity
$d - 2\ln S$ agrees with $Q_{\mathrm{DMAP}}$ within $0.4$--$7\%$,
while the significances $\sim\!\sigma_S$ and $\sim\!\sigma$ agree
within $0.6\sigma$. The two independent measurements of the number of
degrees of freedom are also broadly consistent. For example, for
$\Lambda$CDM one has $k_{\mathrm{eff}} = 2.97$ versus $d = 3.35$. The
Bayesian dimensionality difference $d$ in the leading four-parameter
models ($\Lambda$CDM and THDE ($C=0.7$) with varying $G$) is
$d \approx 1.6$--$1.9$. Thus, the residual inconsistency occupies
about two effective degrees of freedom of the full parameter space. For
the full six-parameter THDE model, $d \approx 3.2$.

Thus, suspiciousness serves as an independent Bayesian check of the
conclusion that there is strong tension in the $\Lambda$CDM model. It
also allows us to test additionally whether the reduction of tension in
the extended models is real or is partly caused by the broadening of
the posteriors and prior effects.

\section{Conclusion}

We have considered cosmological models with a time-varying gravitational constant $G$ in which the dark energy component is described by Tsallis holographic dark energy. Our analysis shows that the time-dependence of the gravitational constant significantly alters the cosmological evolution both in the past and in the future. In these models, phantomization can occur when the effective equation-of-state parameter becomes less than $-1$. However, this phantomization does not necessarily imply a future singularity. For example, for $C=1$ and $\gamma=1$, we find that $w \rightarrow -1$ as $t \rightarrow \infty$ in the standard Friedmann cosmology. For $G' < 0$, phantomization occurs in the future for realistic model parameters, yet no future singularity arises. For $C > 1$, the Hubble parameter decreases with time, even though $w < -1$ for certain parameter choices.

The main result of our investigation is that these models are capable of describing observational data from various probes. Nevertheless, the considered models face certain challenges. Observational data for the Hubble parameter at various redshifts favor for all $C$ a decreasing gravitational constant in the past ($G' < 0$). For supernova data at $C<1$ we have better description for $\alpha_G\sim - 0.1$ but for $C\geq 1$ situation changes in favor to $\alpha_G\sim 0.1$. For baryon acoustic oscillation (BAO) distance indicators we have that the optimal value is $\alpha_G \approx 0$ for $C<0$ and $\alpha_G$ prefers lower bound for $C>1$.

It is noteworthy that a separate analysis of individual datasets showed that THDE models offer no significant advantages over the $\Lambda$CDM model. However, when Pantheon+, $H(z)$, and BAO data are considered jointly, the situation changes. Firstly, from a combined analysis of the various observational datasets, we concluded that models with different values of the parameter $C$ exhibit distinct behavior. For $C \geq 1$, the agreement with observational data is significantly worse than for the $\Lambda$CDM model. Secondly, we found that for $G'/G = 0.1$ and $C\approx 0.6$, this model describes all observational probes better than the $\Lambda$CDM model with varying gravitational constant and especially compared to the standard model.

We have performed a detailed statistical comparison of the Tsallis Holographic Dark Energy (THDE) model with a varying gravitational constant $G$ against the standard $\Lambda$CDM model, using combined data from Type Ia supernovae (Pantheon+), baryon acoustic oscillations (BAO), and Hubble parameter $H(z)$ measurements.

Based on the analysis presented in Sections III and IV, the following key conclusions can be drawn:
\begin{enumerate}
\item \textbf{Preference for models with varying $G$:} Statistical metrics such as the Deviance Information Criterion (DIC) and the Bayes factor unambiguously indicate that models with a varying gravitational constant ($\alpha_G \neq 0$) {provide a better description of the observational data than $\Lambda$CDM. For all leading models with varying $G$, we find $\Delta \mathrm{DIC} \lesssim -22$ and $\ln B_{i0} \ge +10$, Of course quantitative strength of the statistical preference can depend on the adopted prior ranges in any case.}

\item \textbf{Crucial role of the parameter $C$:} The best agreement with the data is achieved for the THDE model with $C < 1$ (specifically, $C \approx 0.6$). For this parameter range, we observe a substantial improvement in the fit quality: for example if $C=0.7$ then $\chi^2_{\min} \approx 634.5$, which is about 30 units better than that of the standard $\Lambda$CDM model ($\chi^2_{\min} \approx 664.9$). For $C \ge 1$, the quality of the fit deteriorates significantly, and the model becomes less favoured than $\Lambda$CDM.

\item \textbf{Tension analysis:} Our analysis using the Index of Inconsistency (IOI) and the $Q_{\mathrm{DMAP}}$ metric shows that the tension between different data sets (particularly between $H(z)$ and SNe Ia) is partially alleviated in models with varying $G$. However, it is important to note that for the THDE model with $G = \mathrm{const}$, the reduction in tension indicated by the IOI metric is an artifact caused by the broadening of the posterior distributions, and is not confirmed by the more robust $Q_{\mathrm{DMAP}}$ metric. In contrast, for the THDE model with varying $G$ and $C < 1$, we observe a genuine reduction in tension, confirmed both by the decrease in $Q_{\mathrm{DMAP}}$ (from 62.8 to 32.4, corresponding to a significance reduction from $\sim 7.4\sigma$ to $\sim 5.3\sigma$) and by the Bayesian suspiciousness metric ($\ln S$).

\item \textbf{Interpretation and limitations:} Despite the significant improvement in the statistical description of the data, the THDE models with varying $G$ face certain challenges. For instance, different individual data sets (SNe Ia and $H(z)$) prefer opposite signs of $\alpha_G$, which is a source of residual tension. Furthermore, in most of the best-fit models, the optimal value of $\alpha_G$ reaches the upper boundary of the prior interval ($\alpha_G \to 0.1$), indicating that the true minimum may lie beyond the considered range, and that our estimates are therefore conservative.

{Importantly, more stringent constraints on the possible time variation of the gravitational constant may emerge in the future. The optimal value of the \(\alpha _{G}\) parameter found by us lies at the edge of current observational limits. If the allowed threshold decreases, it will also reduce the advantage of the proposed model.}

\end{enumerate}

{It is worth noting that the constant-$\alpha_G$ approximation used in our investigation should be understood strictly within the context of a low-energy effective model rather than a fundamental description valid throughout cosmic history. In a more general dark energy scenario, the 
parameter $\dot{G}/G$ is expected to be a dynamical variable that could 
dynamically screen or diminish at $z \to 0$, ensuring compatibility with 
the highly restrictive local constraints. The 
non-zero preferred value of $\alpha_G$ in this work represents a 
redshift-averaged effective contribution that parameterizes the overall 
evolutionary trend in the late universe.}

In summary, we conclude that the combination of the holographic approach (Tsallis HDE) with a varying gravitational constant represents a phenomenologically rich and statistically well-motivated alternative to the $\Lambda$CDM model. The case $C < 1$ is particularly attractive, as it not only provides a better fit to the data but also shows a tendency to alleviate the current tensions in cosmology. Nevertheless, further investigation, including analysis of data at higher redshifts and consideration of possible interactions between dark energy and dark matter, is required for a definitive confirmation of this scenario.

\section*{Acknowledgments}

This work was supported by Ministry of Education and Science (Russia), project 075-02-2026-1311.

\end{document}